\documentclass[twocolumn]{aastex62}

\usepackage{upgreek}
\usepackage{amsmath}

\newcommand{\rone}{FRB~20121102A}
\newcommand{\rthree}{FRB~20180916B}
\newcommand{\periodicitypaper}{PR3}
\newcommand{\msun}{M_\odot}

\newcommand{\mcgillphysics}{Department of Physics, McGill University, 3600 rue University, Montr\'eal, QC H3A 2T8, Canada}
\newcommand{\msi}{McGill Space Institute, McGill University, 3550 rue University, Montr\'eal, QC H3A 2A7, Canada}
\newcommand{\astron}{ASTRON, The Netherlands Institute for Radio Astronomy, Oude Hoogeveensedijk 4, 7991 PD Dwingeloo, The Netherlands}
\newcommand{\api}{Anton Pannekoek Institute for Astronomy, University of Amsterdam, Science Park 904, 1098 XH Amsterdam, The Netherlands}
\newcommand{\oxford}{University of Oxford, Sub-Department of Astrophysics, Denys Wilkinson Building, Keble Road, Oxford, OX1 3RH, United Kingdom}
\newcommand{\wvuphysics}{Department of Physics and Astronomy, West Virginia University, P.O. Box 6315, Morgantown, WV 26506, USA}
\newcommand{\wvugws}{Center for Gravitational Waves and Cosmology, West Virginia University, Chestnut Ridge Research Building, Morgantown, WV 26505, USA}
\newcommand{\uoftphysics}{Department of Physics, University of Toronto, 60 St. George Street, Toronto, ON M5S 1A7, Canada}
\newcommand{\cita}{Canadian Institute for Theoretical Astrophysics, 60 St. George Street, Toronto, ON M5S 3H8, Canada}
\newcommand{\dunlapinstitute}{Dunlap Institute for Astronomy \& Astrophysics, University of Toronto, 50 St. George Street, Toronto, ON M5S 3H4, Canada}
\newcommand{\dunlapdep}{David A. Dunlap Department of Astronomy \& Astrophysics, University of Toronto, 50 St. George Street, Toronto, ON M5S 3H4, Canada}
\newcommand{\mpifr}{Max Planck Institute for Radio Astronomy, Auf dem H\"{u}gel 69, 53121 Bonn, Germany}
\newcommand{\vlbi}{Joint Institute for VLBI ERIC, Oude Hoogeveensedijk 4, 7991PD Dwingeloo, The Netherlands}
\newcommand{\onsala}{Department of Space, Earth and Environment, Chalmers University of Technology, Onsala Space Observatory, 439 92, Onsala, Sweden}
\newcommand{\asc}{Astro Space Centre, Lebedev Physical Institute, Russian Academy of Sciences, Profsoyuznaya Str. 84/32, Moscow 117997, Russia}
\newcommand{\mitkavli}{MIT Kavli Institute for Astrophysics and Space Research, Massachusetts Institute of Technology, 77 Massachusetts Ave, Cambridge, MA 02139, USA}
\newcommand{\mitphysics}{Department of Physics, Massachusetts Institute of Technology, 77 Massachusetts Ave, Cambridge, MA 02139, USA}
\newcommand{\ubc}{Dept. of Physics and Astronomy, 6224 Agricultural Road, Vancouver, BC V6T 1Z1 Canada}
\newcommand{\sidrat}{Sidrat Research, PO Box 73527 RPO Wychwood, Toronto, Ontario, M6C 4A7, Canada}
\newcommand{\perimeter}{Perimeter Institute for Theoretical Physics, 31 Caroline Street N, Waterloo ON N2L 2Y5 Canada}
\newcommand{\tata}{Department of Astronomy and Astrophysics, Tata Institute of Fundamental Research, Mumbai, 400005, India}
\newcommand{\ncra}{National Centre for Radio Astrophysics, Post Bag 3, Ganeshkhind, Pune, 411007, India}

\def\lapp{\ifmmode\stackrel{<}{_{\sim}}\else$\stackrel{<}{_{\sim}}$\fi}
\def\gapp{\ifmmode\stackrel{>}{_{\sim}}\else$\stackrel{>}{_{\sim}}$\fi}

\begin{document}

\title{LOFAR Detection of 110--188 MHz Emission and Frequency-Dependent Activity from \rthree}

\correspondingauthor{Ziggy Pleunis}
\email{ziggy.pleunis@physics.mcgill.ca}

\author[0000-0002-4795-697X]{Z.~Pleunis}
\affiliation{\mcgillphysics}
\affiliation{\msi}

\author[0000-0002-2551-7554]{D.~Michilli}
\affiliation{\mcgillphysics}
\affiliation{\msi}

\author[0000-0002-1429-9010]{C.~G.~Bassa}
\affiliation{\astron}

\author[0000-0003-2317-1446]{J.~W.~T.~Hessels}
\affiliation{\astron}
\affiliation{\api}

\author[0000-0002-9225-9428]{A.~Naidu}
\affiliation{\oxford}

\author[0000-0001-5908-3152]{B.~C.~Andersen}
\affiliation{\mcgillphysics}
\affiliation{\msi}

\author[0000-0002-3426-7606]{P. Chawla}
\affiliation{\mcgillphysics}
\affiliation{\msi}

\author[0000-0001-8384-5049]{E. Fonseca}
\affiliation{\mcgillphysics}
\affiliation{\msi}
\affiliation{\wvuphysics}
\affiliation{\wvugws}

\author[0000-0002-1836-0771]{A.~Gopinath}
\affiliation{\astron}
\affiliation{\api}

\author[0000-0001-9345-0307]{V.~M.~Kaspi}
\affiliation{\mcgillphysics}
\affiliation{\msi}

\author[0000-0001-8864-7471]{V.~I.~Kondratiev}
\affiliation{\astron}
\affiliation{\asc}

\author[0000-0001-7931-0607]{D.~Z.~Li}
\affiliation{\uoftphysics}
\affiliation{\cita}

\author[0000-0002-3615-3514]{M.~Bhardwaj}
\affiliation{\mcgillphysics}
\affiliation{\msi}

\author[0000-0001-8537-9299]{P.~J.~Boyle}
\affiliation{\mcgillphysics}
\affiliation{\msi}

\author[0000-0002-1800-8233]{C.~Brar}
\affiliation{\mcgillphysics}
\affiliation{\msi}

\author[0000-0003-2047-5276]{T.~Cassanelli}
\affiliation{\dunlapinstitute}
\affiliation{\dunlapdep}

\author[0000-0001-5765-0619]{Y.~Gupta}
\affiliation{\ncra}

\author[0000-0003-3059-6223]{A.~Josephy}
\affiliation{\mcgillphysics}
\affiliation{\msi}

\author[0000-0002-5307-2919]{R.~Karuppusamy}
\affiliation{\mpifr}

\author[0000-0002-5575-2774]{A.~Keimpema}
\affiliation{\vlbi}

\author[0000-0001-6664-8668]{F. Kirsten}
\affiliation{\onsala}

\author[0000-0002-4209-7408]{C. Leung}
\affiliation{\mitkavli}
\affiliation{\mitphysics}

\author[0000-0001-9814-2354]{B.~Marcote}
\affiliation{\vlbi}

\author[0000-0002-4279-6946]{K.~Masui}
\affiliation{\mitkavli}
\affiliation{\mitphysics}

\author[0000-0001-7348-6900]{R.~Mckinven}
\affiliation{\dunlapinstitute}
\affiliation{\dunlapdep}

\author[0000-0001-8845-1225]{B.~W.~Meyers}
\affiliation{\ubc}

\author[0000-0002-3616-5160]{C.~Ng}
\affiliation{\dunlapinstitute}

\author[0000-0003-0510-0740]{K.~Nimmo}
\affiliation{\astron}
\affiliation{\api}

\author[0000-0002-5195-335X]{Z. Paragi}
\affiliation{\vlbi}

\author[0000-0003-1842-6096]{M.~Rahman}
\affiliation{\dunlapinstitute}
\affiliation{\sidrat}

\author[0000-0002-7374-7119]{P.~Scholz}
\affiliation{\dunlapinstitute}

\author[0000-0002-6823-2073]{K.~Shin}
\affiliation{\mitkavli}
\affiliation{\mitphysics}

\author{K.~M.~Smith}
\affiliation{\perimeter}

\author[0000-0001-9784-8670]{I.~H.~Stairs}
\affiliation{\ubc}

\author[0000-0003-2548-2926]{S.~P.~Tendulkar}
\affiliation{\tata}
\affiliation{\ncra}

\begin{abstract}
\rthree\ is a well-studied repeating fast radio burst source. Its proximity ($\sim150$\,Mpc), along with detailed studies of the bursts, have revealed many clues about its nature --- including a 16.3-day periodicity in its activity.  Here we report on the detection of 18 bursts using LOFAR at 110--188\,MHz, by far the lowest-frequency detections of any FRB to date.
Some bursts are seen down to the lowest-observed frequency of 110\,MHz, suggesting that their spectra extend even lower.  These observations provide an order-of-magnitude stronger constraint on the optical depth due to free-free absorption in the source's local environment.  The absence of circular polarization and nearly flat polarization angle curves are consistent with burst properties seen at $300-1700$\,MHz.  Compared with higher frequencies, the larger burst widths ($\sim40-160$\,ms at 150\,MHz) and lower linear polarization fractions are likely due to scattering.  We find $\sim2-3$\,rad~m$^{-2}$ variations in the Faraday rotation measure that may be correlated with the activity cycle of the source. 
We compare the LOFAR burst arrival times to those of 38 previously published and 22 newly detected bursts from the uGMRT ($200-450$\,MHz) and CHIME/FRB ($400-800$\,MHz).  Simultaneous observations show 5 CHIME/FRB bursts when no emission is detected by LOFAR.  We find that the burst activity is systematically delayed towards lower frequencies by about three days from 600\,MHz to 150\,MHz.  We discuss these results in the context of a model in which \rthree\ is an interacting binary system featuring a neutron star and high-mass stellar companion.
\end{abstract}

\keywords{Radio transient sources (2008); High energy astrophysics (739); Neutron stars (1108)}

\section{Introduction} \label{sec:intro}

The discovery of radio pulsars \citep{hewish1968} using a low-frequency dipole array (81.5\,MHz) established the existence of neutron stars, and demonstrated that short-duration, coherent radio pulses can be the sirens of extreme astrophysical environments and events. The prediction of coherent radio bursts from other extreme astrophysical settings and events \citep[e.g.,][]{colgate1971} inspired early searches for fast radio transients using archival pulsar survey data \citep[e.g.,][]{phinney1979}. The discovery of the ``Lorimer Burst'' \citep{lor07}, and other bursts with dispersion time delays that place them outside of our Galaxy \citep{tho13}, in archival Parkes pulsar survey data led to the establishment of a population of fast radio bursts (FRBs).

FRBs are sub-second radio flashes that can be detected over extragalactic distances (see \citealt{petroff2019} and \citealt{cordes2019} for recent reviews).  Their physical origin is as yet unclear, but dozens of models have been proposed \citep[see][for a catalog\footnote{\url{https://frbtheorycat.org/}} of theories]{pla18}.  Both repeating \citep{spi16} and apparently non-repeating \citep{petroff2015,shannon2018} FRBs have been detected, and could potentially be created by physically distinct sources or emission mechanisms \citep{chi19_8repeaters,fon20}.  Because of their short duration and high brightness temperature, many models have invoked compact objects as the source of FRBs. The recent discovery of an extremely luminous radio burst from the Galactic magnetar SGR\,J1935+2154 \citep{sgr_chime,sgr_stare} strengthens the case for FRB models that invoke a similar type of source.  In any case, SGR\,J1935+2154 has demonstrated that neutron stars can produce millisecond-duration radio flashes spanning over seven orders-of-magnitude in apparent luminosity \citep{kirsten2020}.

The lack of prompt optical, X-ray or gamma-ray counterparts to FRBs\footnote{Though note that the Galactic event from SGR 1935+2154 was accompanied by a hard X-ray burst \citep{mereghetti2020}.} \citep{scholz17,scholz2020,hardy2017} underscores the need to extract as many useful constraints as possible from the properties of the radio bursts themselves.  Fortunately, detailed spectro-temporal and polarimetric studies of FRBs --- using raw voltage data where possible --- provide important insights into the emission mechanism and local environment \citep[e.g.,][]{farah18,cho2020,day2020,nimmo2020}.
These studies reveal, e.g, that viable emission mechanisms must account for a wide dynamic range of timescales within and between bursts: from a few microseconds up to several milliseconds \citep{nimmo2020}.  The frequency drifts of sub-bursts \citep[i.e., the `sad trombone' effect;][]{hessels19} appear to be a common feature of repeaters \citep{chi19_r2}.  While some repeating FRBs show remarkably similar polarimetric properties --- e.g., \rone\ and \rthree\ show flat polarization position angle within and {\it between} bursts \citep{michilli18,nimmo2020} --- the repeating FRB~180301 shows diverse polarization angle swings between bursts \citep{luo2020}. A larger sample is needed for confirmation, but these spectro-temporal and polarimetric characteristics may indicate a different physical origin or emission mechanism compared to apparent non-repeaters.

Along with \rone, \rthree\, discovered by the Canadian Hydrogen Intensity Mapping Experiment FRB backend \citep[CHIME/FRB;][]{chi19_8repeaters}, is the best-characterized repeating FRB.  It has recently been shown that the burst rate of \rthree\ varies with a period of $16.35\pm0.15$\,days \citep[][hereafter PR3]{r3_period2020}.   \rone\ may also have a $\sim160$-day period to its activity \citep{rajwade2020,cruces2021}.  These activity periods might reflect an orbital period \citep{iokazhang2020, lbg2020, zhanggao2020, popov2020}, rotational period \citep{beniamini2020}, or precession period \citep{levin2020, sobyanin2020, yangzou2020, zanazzi2020}.

\rthree\ is located in a spiral galaxy at a luminosity distance of $D_{\rm L} = 149$\,Mpc \citep{marcote2020}.  This distance makes \rthree\ by far the closest known FRB source with a precise localization; in fact, it is also the most precisely localized FRB to date.  The 2.3-mas localization provided by the European Very-long-baseline interferometry Network \citep[EVN;][]{marcote2020}, coupled with $60-90$-mas imaging from the \emph{Hubble Space Telescope} (\emph{HST}), demonstrates that \rthree\ is close to, but still offset by $\sim250$\,pc from the nearest knot of star formation in the host galaxy \citep{tendulkar2020}.  This suggests that \rthree\ may be too old (100\,kyr$-$10\,Myr) to host an active magnetar.  Rather, \citet{tendulkar2020} argue that \rthree\ may be a high-mass X-ray binary (HMXB) where interaction between the companion wind and neutron-star magnetosphere produces FRBs.

To date, FRBs have been detected from  radio frequencies of 300\,MHz \citep{chawla2020,pilia2020,pck+20} up to 8\,GHz \citep{gajjar2018}.  \rthree\ has thus far only been detected up to 1.7\,GHz \citep{marcote2020}, and \citet{pearlman2020} demonstrate that it is either less active or fainter at higher frequencies ($\gtrsim2$\,GHz).  In simultaneous Low-Frequency Array (LOFAR) and Green Bank Telescope (GBT) observations, \citet{chawla2020} demonstrate burst detections at $300-400$\,MHz, while no emission is seen contemporaneously at $110-188$\,MHz.  Likewise, \citet{pearlman2020} use Deep Space Network 70-m dish observations to demonstrate that no emission is detected at 2.3\,GHz or 8.4\,GHz at the time of a bright burst seen by CHIME/FRB from $600-800$\,MHz.  Such narrow-band emission appears to be a characteristic of repeating FRBs \citep[e.g.,][]{kumar2021}, and has also been well demonstrated for \rone\ \citep{gourdji2019,majid2020}.

\citet{fa19a,fa19b} present FRB candidates detected at 111\,MHz with the Large Phased Antenna of the Lebedev Physical Institute. We consider it difficult to establish an unambiguous astrophysical origin for these signals due to the narrow receiver bandwidth used (2.5\,MHz over 6 frequency channels; complicating the confirmation of dispersion delay proportional to $\nu^{-2}$), the large number of trials in their blind search and the low S/N of the claimed events.

Detecting FRBs at very low radio frequencies ($<300$\,MHz) is challenging: sky background temperature ($T_{\rm sky}$), increase in the effect of scatter broadening in the intervening ionized medium, and uncorrected intra-channel dispersive smearing can all reduce the observed signal-to-noise ratio (S/N).  Nonetheless, low-frequency searches can provide strong constraints on the emission mechanism and local environment of an FRB, e.g., by quantifying the influence of free-free absorption, where the optical depth scales roughly quadratically with radio frequency, $\tau_{\rm ff}\propto\nu^{-2.1}$.  Low-frequency FRB searches are thus well motivated scientifically, despite the increased observational challenges.  More than 50 years after the seminal discovery of radio pulsars, we are now using broad-band, low-frequency dipole arrays with state-of-the-art digital backends to search for extragalactic radio bursts at frequencies below 300\,MHz.

And yet, to date no FRBs have been clearly detected below 300\,MHz --- despite both simultaneous, multi-frequency, targeted \citep{law2017,sokolowski2018,hou19} and blind, wide-field \citep{coe14,kar15,tingay2015,row16,scb+19} searches using the Low-Frequency Array (LOFAR), Murchison Wide-field Array (MWA) and Long-Wavelength Array (LWA).

Here we present LOFAR\footnote{Half of these LOFAR observations (56\,hours out of a total 112\,hours) are also reported on by an independent group in parallel \citep{pcl+20}.} high-band antenna (HBA; $110-188$\,MHz), upgraded Giant Metre Wavelength Radio Telescope (uGMRT; $200-450$\,MHz) and CHIME/FRB ($400-800$\,MHz) observations of \rthree. These observations achieve, by far, the lowest-frequency FRB detections to date, and provide an unprecedented data set to investigate whether the observed activity of the source systematically depends on radio frequency.  We present the observations in \S2 and \S3, and discuss their implications for FRB source and emission models in \S4. We conclude in \S5.

\section{Observations} \label{sec:obs}

\begin{figure*}
  \centering
  \includegraphics[width=\textwidth]{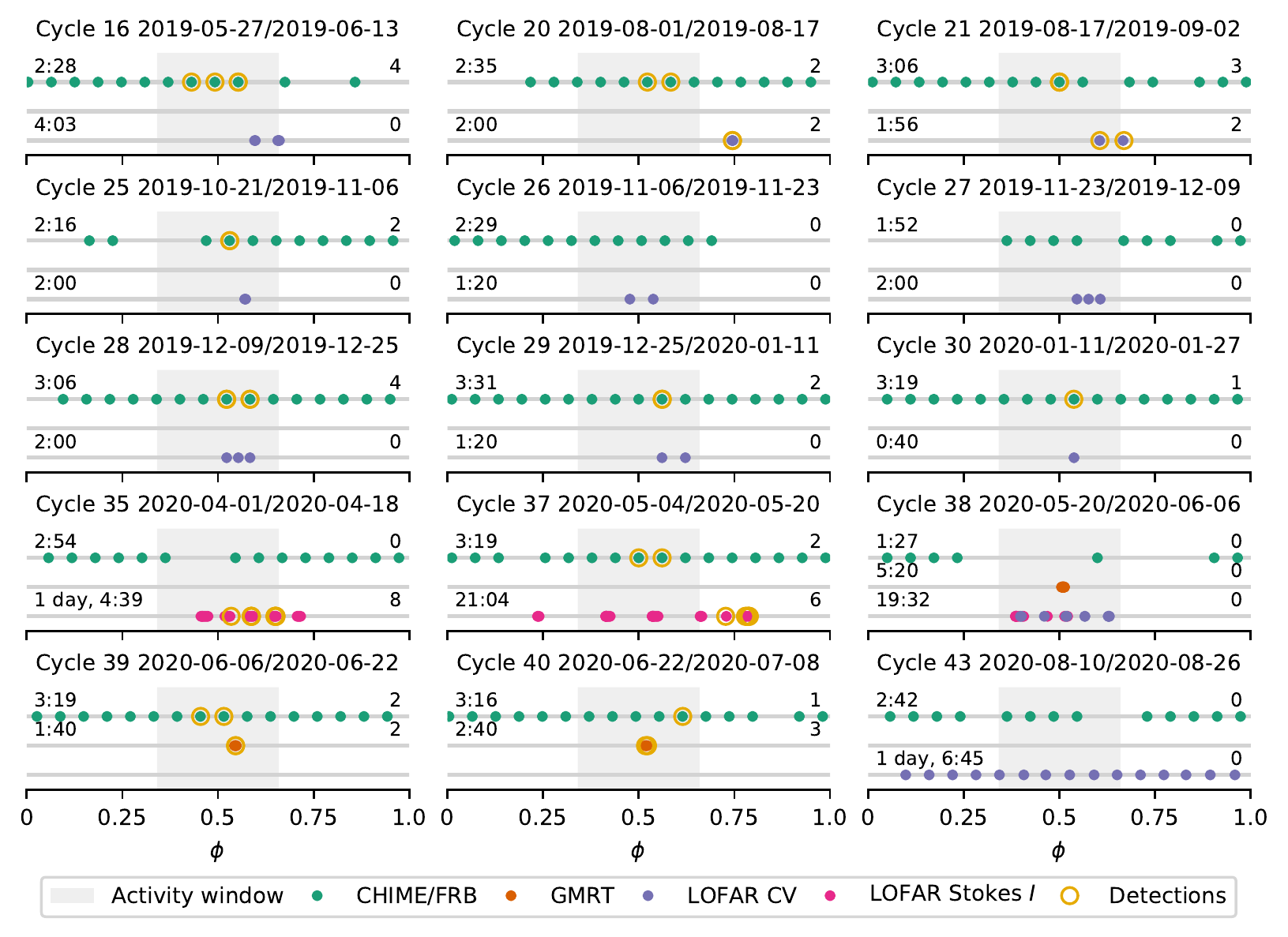}
\caption{
    Summary of per-cycle observations for CHIME/FRB (top row), uGMRT (middle row) and LOFAR (bottom row), as a function of activity phase. In each row the text on the left shows the total observing time (in hours and minutes) and the text on the right shows the total number of bursts detected in the cycle, for the respective telescopes. Only cycles in which uGMRT and LOFAR were observing are shown. Note that the new CHIME/FRB detections reported here are in Cycle 32 and later and that not all of them are shown in this figure. Cycle 1 is the first cycle in which CHIME/FRB detected a burst from the source, with $\phi_0 = 58369.40$ MJD, such that $\phi=0.5$ is the mean of the folded phases of the CHIME/FRB bursts. ``CV'' stands for complex voltage data.
\label{fig:obs}}
\end{figure*}

Figure~\ref{fig:obs} presents an overview of the LOFAR, uGMRT and CHIME/FRB observations taken in 15 different, and not necessarily contiguous, 16.3-day cycles of activity of \rthree\ (as determined from CHIME/FRB detections). Throughout this paper, a dispersion constant of $k_\mathrm{DM}=(2.41\times10^{-4})^{-1}$\,MHz$^2$\,cm$^3$\,pc$^{-1}$\,s is used, following the pulsar convention (\citealt{mt72}; see discussion in \citealt{kul20}).

\begin{figure*}
  \centering
   \includegraphics[width=0.95\textwidth]{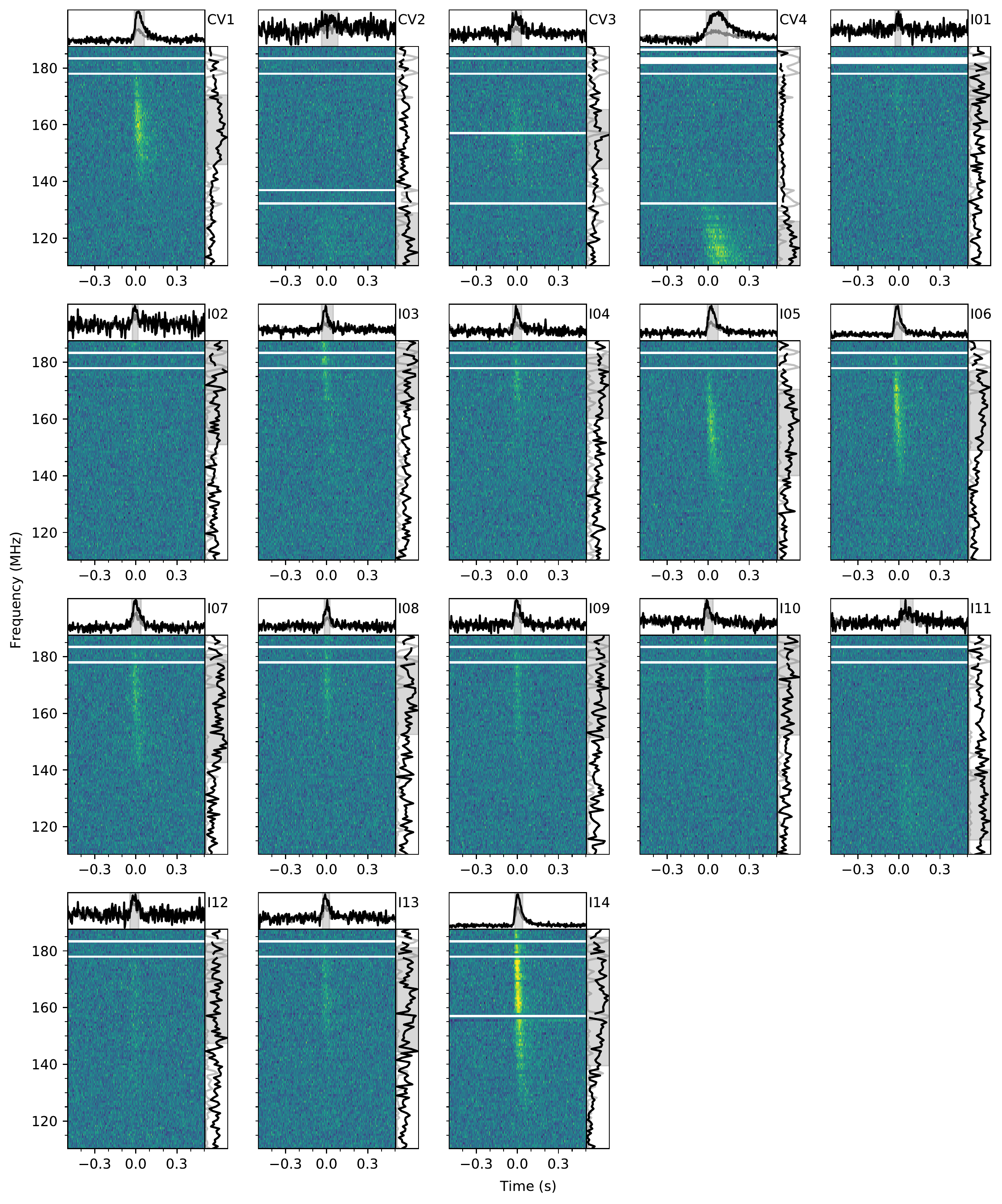}
    \caption{Dynamic spectra of bursts detected with LOFAR. All bursts
    are dedispersed to $\mathrm{DM}=348.772$\,pc\,cm$^{-3}$, the best-fit DM from
    \citet{nimmo2020}. These dynamic spectra have been averaged to a time resolution
    of 3.93\,ms and a frequency resolution of 0.781\,MHz, and the average bandpass of the off-pulse region has been subtracted. The color scale is set to be the same in all panels. The horizontal white bars indicate parts of the spectra where radio frequency interference was masked. Time-averaged spectra are shown at the
    right-hand side of each panel (black), as well as the fraction of averaged
    frequency and time points that were masked to excise radio frequency interference (light grey). The grey band in these panels denotes the burst FWHM in frequency. The top
    panel shows the frequency-averaged pulse profile over the spectral envelope of the burst (black) and the entire band ($110-188$\,MHz; grey), with the burst FWHM
    denoted by the grey band. Events labeled ``CVn'' are from complex voltage data, while those labeled ``In'' are from total intensity data.}  
  \label{fig:lofar_waterfalls}
\end{figure*}

\subsection{LOFAR}
\label{sec:obs:lofar}

\rthree\ was observed with LOFAR \citep{hwg+13} using its beam-formed modes \citep{sha+11} for 112\,hours on 128
occasions between 2019\,June\,6 and 2020\,August\,26. The
\textsc{Cobalt} correlator and beamformer \citep{bmn+18} coherently
combined signals from the HBA antennas of the LOFAR Core
stations to create a tied-array beam pointing to the best-known
position of \rthree. The best CHIME/FRB localization of
$\alpha_\mathrm{J2000}=01^\mathrm{h}57^\mathrm{m}43^\mathrm{s}$,
$\delta_\mathrm{J2000}=+65\degr42\arcmin00\arcsec$ (estimated uncertainty of $\sim2^{\prime}$; derived from baseband data) was used until
2019\,August, while the much more precise position from the EVN localization, $\alpha_\mathrm{J2000}=01^\mathrm{h}58^\mathrm{m}00\fs7502$,
$\delta_\mathrm{J2000}=+65\degr43\arcmin00\farcs3152$ (2.3-mas uncertainty), from
\citet{marcote2020} was used afterwards. The angular separation between the two positions is $2\farcm1$. As all observations used either the
innermost 22 or 20 LOFAR Core stations, the $\sim3\arcmin$ FWHM of the
tied-array beam was slightly offset from the actual celestial position of
\rthree\ in the observations prior to
2019\,August, which led to a factor $\sim2$ lower sensitivity for those early observations. Observations were obtained with the source at altitudes
ranging from 37\degr\ up to culmination at 77\degr, with 86\% of the observations
being obtained at altitudes of 60\degr\ or higher.

The earliest LOFAR observations (up to approximately source activity Cycle 26; see Figure~\ref{fig:obs}) were scheduled in response to CHIME/FRB detections. After the identification of a 16.3-day activity period by CHIME/FRB, observations were scheduled close to the peak in the CHIME/FRB-derived activity window (Cycle 26 and later). The observations in Cycle 43 were intended to broadly cover the full range of activity phases.

For all observations, 400 subbands of 195.3125\,kHz each were
recorded, covering observing frequencies of 110 to 188\,MHz for a
total bandwidth of 78.125\,MHz. For 56\,hours of the observations, the
\textsc{Cobalt} correlator and beamformer generated total-intensity Stokes $I$
filterbank data, with a frequency and time resolution of 3.05\,kHz and
983.04\,$\upmu$s, respectively. Nyquist-sampled, dual-polarization
complex voltages (``CV'' data, with 195.3125\,kHz frequency resolution and
5.12\,$\upmu$s time resolution) were recorded for the other 56\,hours of observations. For the dispersion measure of \rthree\ $\mathrm{DM}=348.772 \pm 0.006$\,pc\,cm$^{-3}$ \citep{nimmo2020} the dispersion
delay from infinite frequency to the top of the observed LOFAR band
(188\,MHz) is 40.9\,s, and the delay within the band, from 188\,MHz down to 110\,MHz, is 78.7\,s.

The CV data allow for coherent dedispersion, a technique that we are employing to search for FRBs and millisecond pulsars using LOFAR and our DRAGNET GPU cluster \citep{bassa2017a,bassa2017b}.  In this paper, we apply coherent dedispersion for burst characterization (see \S\ref{sec:results}), but given the large burst widths, and to allow for a homogeneous burst search of both the Stokes $I$ and the
complex voltage datasets, we chose not to coherently dedisperse the
CV data for the purposes of searching for signals. Instead, the complex voltage data were
channelized, time averaged and the polarizations were squared and summed offline to form
Stokes $I$ filterbanks with the same time and frequency resolution as
the Stokes $I$ filterbanks (3.05\,kHz and
983.04\,$\upmu$s) generated by \textsc{Cobalt} in real time. To reduce their size, these filterbanks were first
averaged in frequency by a factor of 16 with \texttt{digifil}
\citep{sb10}, using incoherent dedispersion to $\mathrm{DM}=350$\,pc\,cm$^{-3}$. With this setup, dispersive smearing
due to incoherent dedispersion at the DM of
\rthree\ varies from 1.3 to 6.7\,ms over
the LOFAR band.  This temporal smearing is negligible compared to the burst widths.  Radio frequency interference (RFI) was identified
using the \texttt{rfifind} tool from the \textsc{Presto} software
\citep{ran01, rem02} and replaced with random noise of the appropriate
mean and standard deviation. Next, dedispersed time series were
generated between DMs of 300 to 400\,pc\,cm$^{-3}$, with steps of
0.1\,pc\,cm$^{-3}$, using the GPU-accelerated \textsc{Dedisp}
dedispersion library \citep{bbbf12}. 

To maximize sensitivity towards
possibly narrow-band radio bursts \citep[motivated by][]{gourdji2019,kumar2021}, these time series were created for
the full observing band (110--188\,MHz), as well as three overlapping
halves (110--149\,MHz, 130--169\,MHz and 149--188\,MHz), and seven overlapping quarters of the band (110--129\,MHz, 120--139\,MHz, etc.). These time series were cross-correlated with top-hat
functions with widths up to 150\,ms, using a GPU-accelerated version of
\texttt{single\_pulse\_search.py} from \textsc{Presto} to search for
bursts. All candidate burst events with $\mathrm{S}/\mathrm{N}>7$ were
visually inspected to distinguish bursts from residual RFI. For the 12 data sets with known bursts, we verified that the burst selection is complete by assessing the burst candidates using the FETCH deep-learning-based classifier \citep{agarwal2020}. All the single pulses identified were grouped using Single-pulse Searcher \citep{sps}, and redundant burst candidates were eliminated before putting them through the FETCH classifier.

\subsection{uGMRT}

The uGMRT \citep{gupta2017} observations of \rthree\ were carried out on three different days:
2020\,May\,29, 2020\,June\,15 and 2020\,July\,1.
The observations were intentionally scheduled close to the peak of the CHIME/FRB-derived activity window, and the telescope was phased-up towards the EVN position of \rthree\ during all observations.
On 2020\,May\,29, 
we used both Band~2 (180--280\,MHz) and Band~3 (250--350\,MHz) 
simultaneously with two phased-array beams employing two sub-arrays: 
one sub-array using 10 of the available 29 central-square antennas at Band~2 (180--280\,MHz), and another sub-array at 
Band~3 (250--500\,MHz) using the remaining 19 antennas. We used the uGMRT Wideband Backend \citep[GWB;][]{reddy17} to record 
coherently dedispersed total-intensity filterbank data with 
the passband split into 2,048 channels with 327.68 $\upmu$s 
sampling time in both phased-array beams. The observations were 
divided into eight 20-minute sessions with a provision to re-phase 
the sub-arrays in between the sessions to account for the temporal 
instrumental gain and ionospheric changes. We 
used the same setup on June 15 with five 20-minute sessions, but 
the Band~2 data were not usable due to strong RFI. On 2020 July 1, we
observed only in Band~3 set to 250--450 MHz using 24 antennas in 
four 40-minute sessions. The coherently dedispersed filterbank data 
were recorded with 81.92-$\upmu$s time resolution and 2,048 frequency channels 
across the 200-MHz band. On all three days we have also recorded 
``ON'' source and ``OFF'' source data on 3C48 for calibration purposes.

To prepare the data for searching, we identified and mitigated 
narrow-band RFI and broad-band time-domain RFI using 
\texttt{gptool} \citep{Susobhanan20}. The single-pulse search 
was carried out using a machine-learning technique based 
on \cite{zhang18}. We have trained our convolutional neural network 
model using archival uGMRT data and simulated CHIME/FRB-like FRBs 
with various burst morphologies. The details of our implementation 
will be published elsewhere. This particular search is tuned for 
200 $\leq$ DM $\leq$ 500 \,pc\,cm$^{-3}$ and is sensitive to 
pulse widths up to 256 ms. The dynamic spectra of candidates were 
visually examined to distinguish astrophysical signals from spurious RFI.

\subsection{CHIME/FRB}
CHIME/FRB searches intensity data from 1,024 stationary synthesized beams for dispersed transients over a $\sim$200 deg$^2$ field-of-view in the 400--800 MHz octave in real time \citep{chi18_overview}. The intensity data have a 0.98304-ms time resolution and 16,384 frequency channels.

The median daily exposure of the experiment to the sky position of \rthree~is 746 s (i.e., the time spent by the source within the FWHM of the synthesized beams at 600 MHz). The transit time of the source through the primary beam of the instrument, however, is much longer ($\sim$40 minutes), albeit with significant variation in sensitivity that is still in the process of being quantified outside of the FWHM of the synthesized beams.

\begin{table*}
  \centering
  \footnotesize
  \caption{Burst parameters. See \S\ref{sec:results} for a description of how parameters were determined. For all bursts, arrival times and burst width $\sigma_\mathrm{t}$ are computed for $\mathrm{DM}=348.772$\,pc\,cm$^{-3}$ \citep{nimmo2020}. 
  \label{tab:lofar_burst_properties}}
  \begin{tabular}{lccccccccc}
    \tableline
    Burst & \multicolumn{2}{c}{Barycentric arrival time (TDB) at $\nu=\infty$} & $\phi$ & $\sigma_\mathrm{t}$\tablenotemark{a} & $\nu_\mathrm{low}$ & $\nu_\mathrm{high}$ & S/N\tablenotemark{b} & Fluence & Peak flux density \\
    & (UTC) & (MJD) & & (ms) & (MHz) & (MHz) & & (Jy ms) & (Jy) \\
    \tableline
    \multicolumn{10}{c}{LOFAR} \\
    \tableline
CV1                  & 2019-08-13T03:26:33.454 & 58708.14344 & 0.74363 & 73(3)  & 133.7 & 182.8 & 28.6 & 148(8) & 4.7(1) \\
CV2\tablenotemark{c} & 2019-08-13T04:01:43.268 & 58708.16786 & 0.74512 & 119(35) & 109.8 & 139.0 & 3.9 & 26(9) & 1.51(9) \\
CV3                  & 2019-08-27T04:53:30.828 & 58722.20383 & 0.60464 & 72(7)  & 133.9 & 175.9 & 11.4 & 49(11) & 2.15(7) \\
CV4                  & 2019-08-28T05:14:19.946 & 58723.21829 & 0.66677 & 158(7)  & 109.8 & 135.6 & 25.5 & 196(22) & 3.6(2) \\
I01\tablenotemark{c} & 2020-04-10T15:21:25.278 & 58949.63988 & 0.53214 & 43(10)  & 146.5 & 188.0 & 5.1 & 27(10) & 2.06(8) \\
I02\tablenotemark{c} & 2020-04-11T12:42:02.337 & 58950.52919 & 0.58660 & 42(8)  & 131.3 & 188.0 & 6.3 & 38(9) & 1.85(9) \\
I03                  & 2020-04-11T12:59:28.444 & 58950.54130 & 0.58734 & 87(8)  & 151.5 & 188.0 & 11.8 & 86(10) & 3.0(2) \\
I04                  & 2020-04-11T14:00:12.520 & 58950.58348 & 0.58993 & 66(6)  & 148.9 & 188.0 & 13.2 & 68(11) & 2.75(8) \\
I05                  & 2020-04-12T12:59:56.684 & 58951.54163 & 0.64860 & 80(4)  & 124.8 & 185.7 & 21.5 & 140(9) & 3.75(7) \\
I06                  & 2020-04-12T13:23:32.453 & 58951.55801 & 0.64960 & 58(2)  & 135.2 & 188.0 & 27.8 & 145(13) & 5.4(1) \\
I07                  & 2020-04-12T14:01:58.736 & 58951.58471 & 0.65124 & 69(4)  & 123.5 & 188.0 & 20.0 & 103(9) & 3.9(1) \\
I08                  & 2020-04-12T14:11:32.715 & 58951.59135 & 0.65164 & 41(4)  & 138.6 & 188.0 & 11.5 & 65(10) & 2.9(1) \\
I09                  & 2020-05-16T11:35:48.666 & 58985.48320 & 0.72708 & 50(5)  & 128.2 & 188.0 & 12.7 & 97(39) & 3.4(6) \\
I10                  & 2020-05-17T07:50:25.499 & 58986.32668 & 0.77873 & 63(7)  & 134.7 & 188.0 & 10.5 & 59(10) & 2.7(1) \\
I11\tablenotemark{c} & 2020-05-17T08:21:41.703 & 58986.34840 & 0.78006 & 93(14) & 109.8 & 152.5 & 7.5 & 43(9) & 1.65(5) \\
I12                  & 2020-05-17T10:58:01.354 & 58986.45696 & 0.78671 & 65(8)  & 129.7 & 188.0 & 10.0 & 53(9) & 2.4(1) \\
I13                  & 2020-05-17T11:03:23.282 & 58986.46069 & 0.78694 & 51(5)  & 126.1 & 188.0 & 12.8 & 66(7) & 3.64(5) \\
I14                  & 2020-05-17T11:51:01.413 & 58986.49377 & 0.78896 & 57(2)  & 116.9 & 188.0 & 38.5 & 308(10) & 10.57(7) \\
    \tableline
    \multicolumn{10}{c}{uGMRT} \\
    \tableline
G1 & 2020-06-15T04:25:47.176 & 59015.18457 & 0.54590 & 17(3) & 270\tablenotemark{d} & 350\tablenotemark{d} & 25.0 & 161(71) & 8(2) \\
G2 & 2020-06-15T04:32:28.540 & 59015.18922 & 0.54619 & 12(4) & 280\tablenotemark{d} & 350\tablenotemark{d} & 8.5 & 26(11) & 2.3(4) \\
G3 & 2020-07-01T01:16:03.761 & 59031.05282 & 0.51763 & 39(7) & 325\tablenotemark{d} & 410\tablenotemark{d} & 6.1 & 33(14) & 1.6(3) \\
G4 & 2020-07-01T03:03:41.718 & 59031.12757 & 0.52220 & 18(3) & 375\tablenotemark{d} & 450\tablenotemark{d} & 12.0 & 27(11) & 2.5(4) \\
G5 & 2020-07-01T03:17:40.549 & 59031.13727 & 0.52280 & 98(8) & 300\tablenotemark{d} & 450\tablenotemark{d} & 16.4 & 178(58) & 3.3(8) \\
    \tableline
    \multicolumn{10}{c}{CHIME/FRB} \\
    \tableline
CF39 & 2020-02-19T23:54:17.856 & 58898.99604 & 0.43087 & 4.8(6) & 418 & 520 & 8.8 & $>$1.8(4) & $>$0.4(2) \\ 
CF40 & 2020-02-20T00:10:50.592 & 58899.00753 & 0.43158 & 5.5(7) & 429 & 526 & 11.8 & $>$1.9(6) & $>$0.3(2) \\ 
CF41 & 2020-02-21T00:10:23.808 & 58900.00722 & 0.49279 & 2.6(2), 2.7(2)& 417 & 469 & 24.7 & $>$5(2) & $>$0.7(3) \\ 
CF42 & 2020-03-24T21:33:36.576 & 58932.89834 & 0.50694 & 3.1(3) & 403 & 460 & 13.0 & $>$1.6(6) & $>$0.4(2) \\ 
CF43 & 2020-04-23T19:35:03.264 & 58962.81601 & 0.33901 & 0.54(3) & 400 & 693 & 12.6 & $>$0.9(5) & $>$1.5(6) \\ 
CF44 & 2020-04-24T19:49:22.944 & 58963.82596 & 0.40085 & 1.0(1) & 545 & 674 & 9.2 & $>$2.0(5) & $>$0.8(3) \\ 
CF45 & 2020-05-12T18:33:52.128 & 58981.77352 & 0.49991 & 0.7(2), 2.9(1) & 690 & 800 & 22.7 & 12(3) & 1.7(6) \\ 
CF46 & 2020-05-13T18:26:35.808 & 58982.76847 & 0.56084 & 3.7(4) & 403 & 469 & 12.2 & 2.2(7) & 0.4(3) \\ 
CF47 & 2020-06-13T16:33:47.232 & 59013.69013 & 0.45439 & 2.7(2) & 400 & 479 & 19.3 & $>$4(1) & $>$0.9(4) \\ 
CF48 & 2020-06-14T16:22:57.504 & 59014.68261 & 0.51516 & 0.72(3), 2.6(4) & 539 & 639 & 15.4 & 14(4) & 3.0(8) \\ 
CF49 & 2020-07-02T15:22:45.120 & 59032.64080 & 0.61487 & 4.2(2) & 407 & 462 & 16.2 & $>$5(1) & $>$0.6(3) \\ 
CF50 & 2020-07-17T13:55:55.200 & 59047.58050 & 0.52973 & 3.7(2) & 400 & 432 & 17.7 & $>$4(1) & $>$0.6(2) \\ 
CF51 & 2020-07-17T14:04:18.048 & 59047.58632 & 0.53009 & 2.5(2) & 443 & 511 & 11.5 & $>$0.8(2) & $>$0.2(2) \\ 
CF52 & 2020-08-01T13:26:53.376 & 59062.56034 & 0.44705 & 2.7(4) & 428 & 462 & 8.3 & $>$0.9(3) & $>$0.3(2) \\ 
CF53 & 2020-09-03T11:12:54.720 & 59095.46730 & 0.46217 & 2.9(3) & 536 & 688 & 11.1 & 5(1) & 1.1(4) \\ 
CF54 & 2020-09-04T10:54:30.528 & 59096.45452 & 0.52263 & 1.19(9) & 707 & 800 & 15.8 & $>$2.5(6) & $>$1.2(4) \\ 
CF55 & 2020-09-19T09:48:59.328 & 59111.40902 & 0.43840 & 6.4(7) & 400 & 447 & 10.9 & $>$1.8(3) & $>$0.3(2) \\ 
    \tableline
  \end{tabular}
\tablenotetext{a}{Burst width (FWHM), for a Gaussian function fitted to the time series.}
\tablenotetext{b}{Band-averaged S/N. Four LOFAR-detected bursts were found in a search of the half- and quarter-bandwidth data segments.}
\tablenotetext{c}{Only detected in a sub-band search.}
\tablenotetext{d}{Estimated by eye because baseline variations due to residual broad-band RFI prohibited fitting a model to the burst spectra.}
\end{table*}

\section{Results}\label{sec:results}

All bursts presented here have been
dedispersed to $\mathrm{DM}=348.772$\,pc~cm$^{-3}$, as determined from aligning
sub-structure in EVN voltage data at 1\,$\upmu$s resolution \citep{nimmo2020}.
We decided against optimizing individual burst DMs, as DM and burst morphology are known to be covariant \citep[especially for the majority of low S/N `smudgy' bursts;][]{gourdji2019} and there is as-of-yet no evidence for DM evolution of \rthree\ (\periodicitypaper).

\subsection{LOFAR}\label{ssec:results_lofar}

A total of 18 bursts were detected using LOFAR: 14 were found in the full-bandwidth data, and another 4 fainter bursts were identified in the time series
generated from the half- and quarter-bandwidth data segments. The dynamic spectra of
these bursts are presented in Figure~\ref{fig:lofar_waterfalls}. Four
of the bursts occurred during complex voltage observations, and thus have available
polarimetric information. These are labelled CV1 through CV4, while
the bursts obtained in total-intensity Stokes~$I$ data are labelled I01 through I14.

Bursts I01 through I08 correspond to the bursts L02 though L09 presented by \citet{pcl+20}. Our analysis did not recover their L01 burst due to periodic baseline variations affecting 2.6\,hr of the LOFAR observations. These variations were the result of a rare temporary clock skew at one of the LOFAR stations where some network packets of this station arrived out of sync at the COBALT beamformer \citep{bmn+18} and were discarded. Hence, the number of stations added coherently in the tied-array beam varied at the $\sim1$\,s beamformer block size of COBALT. The resulting baseline variations increased the noise in the dedispersed timeseries of these observations, placing the signal-to-noise of this burst below our detection threshold.

FETCH positively identified the same set of 18 bursts found through visual inspection, including the four faint bursts found only in the sub-band search, using the full-bandwidth data. The 14 bursts identified by visual inspection in the full-bandwidth data were classified as astrophysical pulse candidates with probabilities greater than 83\% by all 11 available models (labeled a to k) of FETCH. The 3 fainter bursts I01, I02 and I11 were positively identified by at least 9 of the 11 models with probabilities greater than 64\%. The faintest burst CV2 could only be identified by FETCH models e and h, each with a probability more than 93\%. No additional bursts were found in these data sets.

The LOFAR burst properties were determined by fitting Gaussian profiles to the time and
frequency averages of the dynamic spectra to obtain their arrival
times, temporal and spectral widths (FHWM) and emission frequencies
(Table~\ref{tab:lofar_burst_properties}). All observed bursts are
band-limited, with spectral widths (defined as FWHM) ranging from 20 to
50\,MHz, and the majority of the bursts, 15 out of 18, peaking in
brightness above 160\,MHz. The temporal width of the bursts varies
between 40\,ms for bursts peaking in the top of the LOFAR band
($\sim180$\,MHz) to 160\,ms near the bottom of the LOFAR band
($\sim120$\,MHz).

To calibrate the LOFAR detections we have subtracted the mean of an off-burst region in each frequency channel and divided by the standard deviation of the off-burst region. We then convert each channel $i$ from S/N to flux units using the radiometer equation:
\begin{equation}
\Delta S_i = \frac{T_{\mathrm{sys},i}}{G_i \sqrt{n_\mathrm{p} \Delta \nu_i t_\mathrm{s}}},
\end{equation}
where $T_{\mathrm{sys},i}$ is the system temperature (receiver and sky), $G_i$ the gain, $n_\mathrm{p} = 2$ the number of summed polarizations, $\Delta \nu_i$ the channel bandwidth and $t_\mathrm{s}$ the sampling time. For the calculation of $T_\mathrm{sys}$ and $G$ we take into account the number of Core stations used and we correct for the zenith-angle dependence, as described
by \citet{kvh+16}. We calculate these values at the times of the burst detections. In the center of the band ($\sim$149 MHz) $T_\mathrm{sys} \approx 1,090$ K (note that \rthree\ is at a Galactic latitude of $b=3.7^{\circ}$ with $T_\mathrm{sky} \approx 700$ K at 150 MHz) and $G \approx 4.3$\,K~Jy$^{-1}$, on average. We repeat this measurement using six different independent, but adjacent, $\sim 100$-ms off-pulse regions in the same observation as the burst and quote the average (band-averaged) fluence and peak flux of the bursts, as well as the standard deviation across the six samples, in Table~\ref{tab:lofar_burst_properties}. Note that the systematic uncertainty in the flux measurements could be larger than reported in Table~\ref{tab:lofar_burst_properties}, due to unaccounted contributions from the ionosphere and the Galactic plane, or other bright sources in the primary beam of the HBAs. As the systematic errors likely underestimate system noise, the fluence measurements are likely also underestimated.

Even after dedispersion to the best-fit DM from \citet{nimmo2020},
the brightest LOFAR bursts show residual time
delays towards lower observing frequencies, with one burst (CV4)
broadening with decreasing frequency. Due to the absence of visible burst
sub-structure, it is unclear if these effects are due to DM underestimation, multi-path scattering, or the frequency drifts of
unresolved sub-bursts \citep[i.e., the `sad trombone' effect;][]{hessels19}. Deviations from the canonical $\Delta t \propto \nu^{-2}$ dispersion relation are expected at low radio frequencies \citep[see][and references therein]{hassall2012}, and it is also possible that the residual time delays in CV4 result from such deviations.

\begin{figure*}
  \centering
   \includegraphics{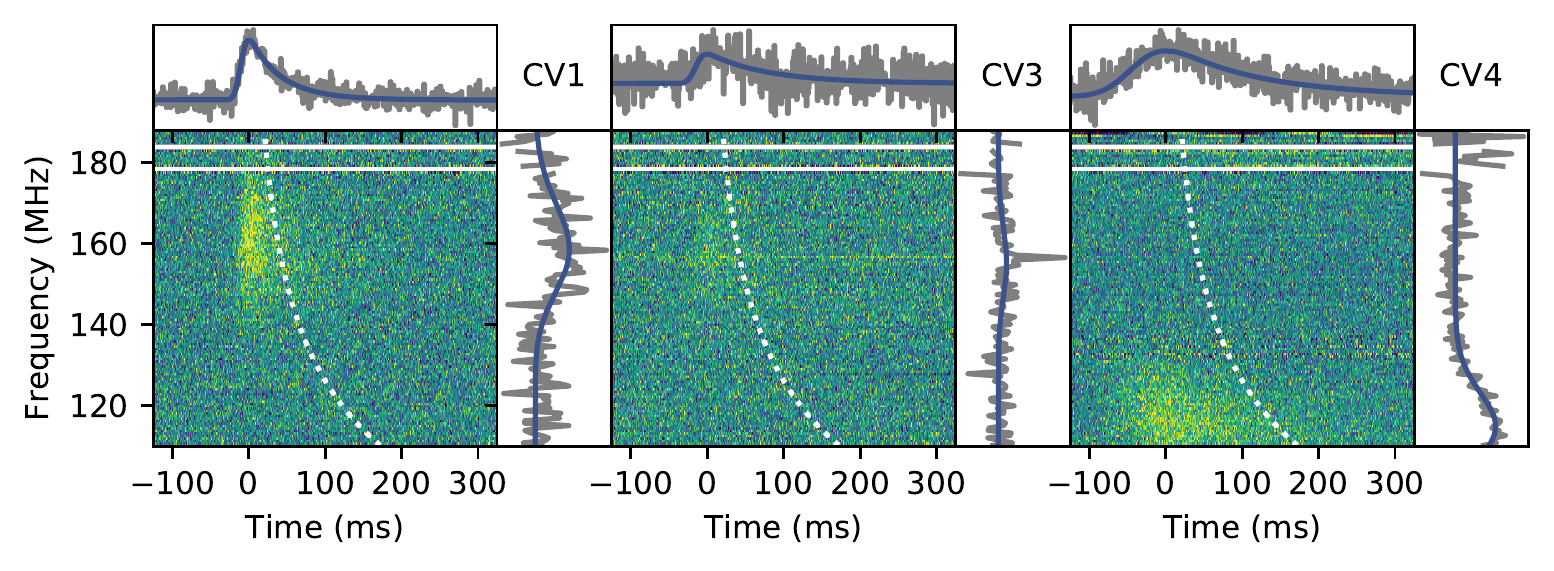}
    \caption{Least-squares burst model fits to LOFAR bursts CV1, CV3 and CV4, after coherent dedispersion. The model (solid blue line; Table~\ref{tab:cv_parameters}) is overlayed on the band-averaged time series (top panel) and the time-averaged spectra (right panel). A fiducial scattering timescale of 50 ms (referenced at 150 MHz) is plotted on top of the dynamic spectra (white dotted line). The horizontal white bars indicate parts of the spectra where radio frequency interference was masked.}
  \label{fig:fitburst}
\end{figure*}

To measure a scattering timescale, we use a least-squares fitting
routine, previously developed for CHIME/FRB bursts \citep[e.g.,][]{josephy2019,chi19_8repeaters}, to model the dynamic spectra of the LOFAR bursts.  We analyze the three brightest
bursts that can be coherently dedispersed: bursts CV1, CV3 and
CV4, where \texttt{dspsr} \citep{sb10} was used to generate coherently dedispersed single pulse profiles. We fixed $\mathrm{DM}=348.772$ pc
cm$^{-3}$ \citep{nimmo2020} and assumed that the bursts consist of only one
component. We referenced the scattering timescale to 150\,MHz and fixed
the scattering index to $-4$. The results are provided in
Table~\ref{tab:cv_parameters} and Figure~\ref{fig:fitburst}. The measured scattering timescale for
the two bursts with highest S/N (CV1 and CV4) is $\sim$50\,ms at 150 MHz.
The other measured timescale, for burst CV3, is more uncertain due to the
low detection S/N and narrow bandwidth of the burst.

\begin{table}
\centering
  \caption{Best-fit intrinsic width, scattering timescale ($\tau_\mathrm{s}$; at 150\,MHz) and rotation measure (RM) for three of the LOFAR bursts.  Only formal fit uncertainties are quoted; this is particularly relevant for the $\tau_\mathrm{s}$ measurement of CV3 (see main text). The RMs are the observed values; they are not corrected for Doppler redshift or ionospheric contribution, although they are corrected for the Earth's motion. The corresponding ionospheric contribution, RM$_\mathrm{iono}$, is reported, as calculated with \texttt{ionFR} (see main text).
  \label{tab:cv_parameters}
  }
  \begin{tabular}{ccccc}
    \tableline
    Burst & Width & $\tau_\mathrm{s}$ & RM$_\text{obs}$ & RM$_\text{iono}$ \\
     & (ms) & (ms) & (rad\,m$^{-2}$) & (rad\,m$^{-2}$) \\
    \tableline
    CV1 & 6.608(1) & 54.142(4) & $-$115.71(3) & 0.30(8) \\
    CV3 & 8.313(4) & 94.55(2) & $-$114.78(9) & 0.34(4) \\
    CV4 & 31.426(3) & 46.692(3) & $-$114.43(4) & 0.39(5) \\
    \tableline
  \end{tabular}
\end{table}

\subsection{Polarization analysis}

We recorded polarization, using LOFAR's orthogonal linear feeds, for bursts CV1--4; three of these bursts are bright enough to perform polarimetric analyses. We corrected for azimuth and elevation-dependent gain using the LOFAR beam model, as implemented within the {\tt dreamBeam} package.\footnote{\url{https://github.com/2baOrNot2ba/dreamBeam}}
No additional polarization calibration, beyond that already implemented to form the individual station and Core tied-array beams \citep{sha+11}, was performed and, thus, slight degradation of the polarization fraction ($\lesssim 5$\%) could arise, e.g., from the thermal expansion of the cables.  The absence of an absolute polarization calibration also prevents us to compare the polarization angle (PA) between different bursts.
In this analysis, the frequency resolution was increased to $\sim12$\,kHz (16 channels synthesized within each 195-kHz subband) to have an intra-channel depolarization smaller than 1\% across the whole band at the value of rotation measure (RM) reported by \citet{chi19_8repeaters}.
We measured the RM value of each burst by using RM synthesis\footnote{\url{https://github.com/CIRADA-Tools/RM-Tools}} \citep{1966MNRAS.133...67B,2005A&A...441.1217B} and a deconvolution algorithm \citep{2009IAUS..259..591H}; the Stokes parameters of each burst have been corrected for the resulting value.

\begin{figure*}
\plotone{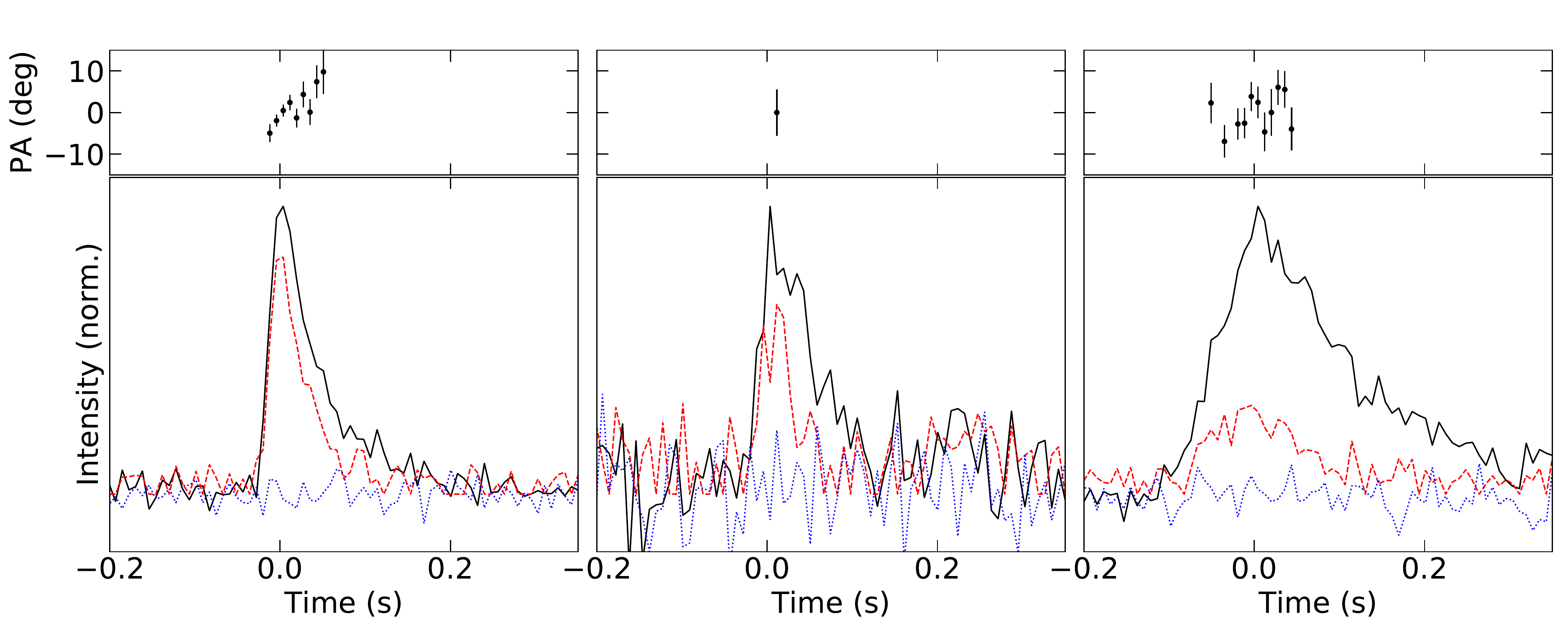}
\caption{
    Polarization profiles obtained for the three LOFAR bursts with available CV data and sufficient S/N. The black curve is the total intensity; the red dashed curve is the linear polarization, after correcting for Faraday rotation; and the blue dotted curve is the circular polarization. 
    For clarity, the profiles have been normalized to all have unitary peak intensities, and are plotted with a time resolution of $7.8125$\,ms.
    The PA curves are reported in the top panel of each profile and are rotated by an arbitrary angle in order to be centered around zero. 
\label{fig:pol_profiles}}
\end{figure*}

We report the polarization profiles for the three bursts in Figure~\ref{fig:pol_profiles}, using the coherently dedispersed pulse profiles from \S\,\ref{ssec:results_lofar}.
The linear polarization fractions of CV1, CV3 and CV4 are roughly 70\%, 60\% and 30\%, respectively --- much lower than the $\sim$ 100\% reported at higher frequencies of $300-1700$\,MHz \citep{chi19_8repeaters,chawla2020,nimmo2020}.
\citet{noutsos2015} performed a long observation of PSR B2217+47 to study the depolarization fraction as a function of the hour angle (HA) for LOFAR. 
They found a depolarization fraction $<10\%$ for $|\text{HA}|<6$\,hours and zenith angle $<50\degr$.
Our observations of \rthree\ are well within this range, with a maximum hour angle of $2.1$\,hours and a maximum zenith angle of $20.3\degr$.
Given also the absence of any visible artifacts in the Faraday dispersion function that could indicate the presence of signal leakage or uncorrected delay between the polarization channels (such as a peak at $\text{RM}=0$\,rad\,m$^{-2}$ or symmetric peaks around $\text{RM}=0$\,rad\,m$^{-2}$) we conclude that the observed depolarization towards lower radio frequencies is predominantly astrophysical in origin.

Part of the depolarization at lower frequencies could therefore be intrinsic to the source emission or be related to extrinsic propagation effects.
In particular, CV4 is observed at the lowest frequencies and has the lowest linear polarization fraction, which is compatible with depolarization due to scattering, as observed in some pulsars \citep[e.g.,][]{noutsos2015,xue19}. 
As at higher frequencies, the circular polarization fraction here is consistent with $0$\%. 
The PA curve is nearly flat at higher frequencies, though subtle variations are seen on short timescales \citep{nimmo2020}. 
While the PA curve of CV4 is consistent with being flat ($\chi_\text{red}^2=1.1$), the PA curve of the brightest burst, CV1, shows a hint of an increase at later times, with a $\chi_\text{red}^2=2.2$ with respect to a straight line.
\citet{day2020} similarly find evidence for a time-varying PA in the wide and potentially scatter-broadened FRB~190608.

\begin{figure*}
\plotone{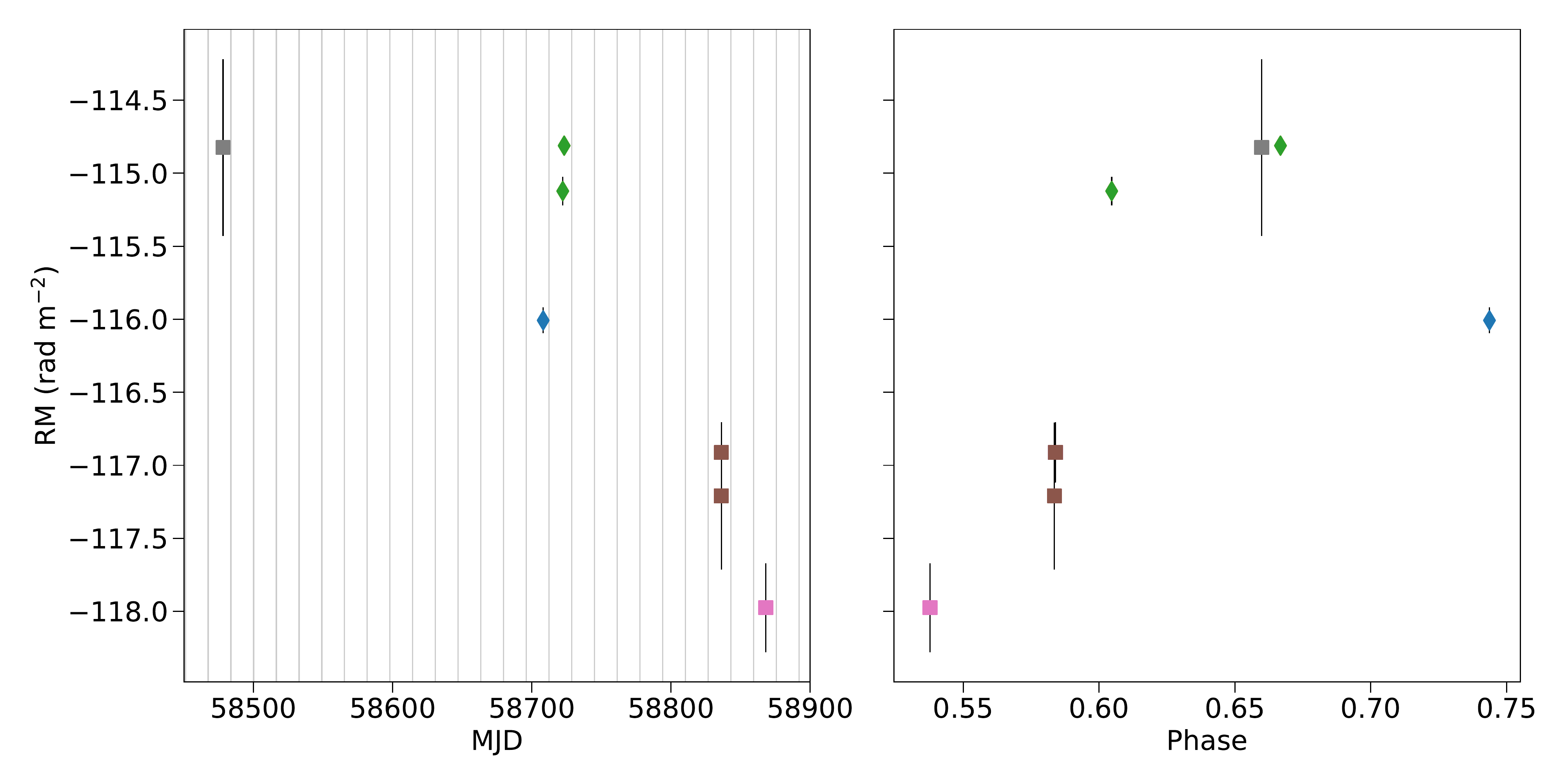}
\caption{
    Measured RMs for \rthree\ from this work (diamond symbols), \citet{chi19_8repeaters} and \citet{chawla2020} (square symbols).  The RMs are corrected for the ionospheric contribution and plotted with 1-$\sigma$ error bars.
    Left: Plot as a function of time (in MJD), where vertical lines represent phase zero of the different activity cycles of the source. 
    Right: Plot as a function of the source activity phase.
    The different colors of the data points represent different source cycles. 
\label{fig:RM}}
\end{figure*}

The lack of refined polarization calibration is not expected to affect the RM values measured by LOFAR \citep[e.g.,][]{sob19}, which are reported in Table~\ref{tab:cv_parameters}. 
We compare the new LOFAR RM values obtained here with other measurements presented by \citet{chi19_8repeaters} and \citet{chawla2020}.
The ionospheric contribution to the observed burst RMs was determined using \texttt{ionFR}\footnote{\url{https://github.com/csobey/ionFR}} \citep{sot13}, which utilizes data from the International Geomagnetic Reference Field and IONosphere map EXchange (IONEX) global ionospheric maps.
The resulting values are shown in Figure~\ref{fig:RM} as a function of date and source activity phase, as measured by \periodicitypaper\ and refined in \textsection\ref{sec:periodic_activity}. 
There is a hint of RM values changing systematically as a function of activity phase, as measured by different instruments in different activity cycles. However, the RM variations may simply be stochastic; more detections are needed to investigate this further.

\subsection{uGMRT}\label{ssec:results_gmrt}

\begin{figure*}
  \centering
    \includegraphics[width=0.95\textwidth]{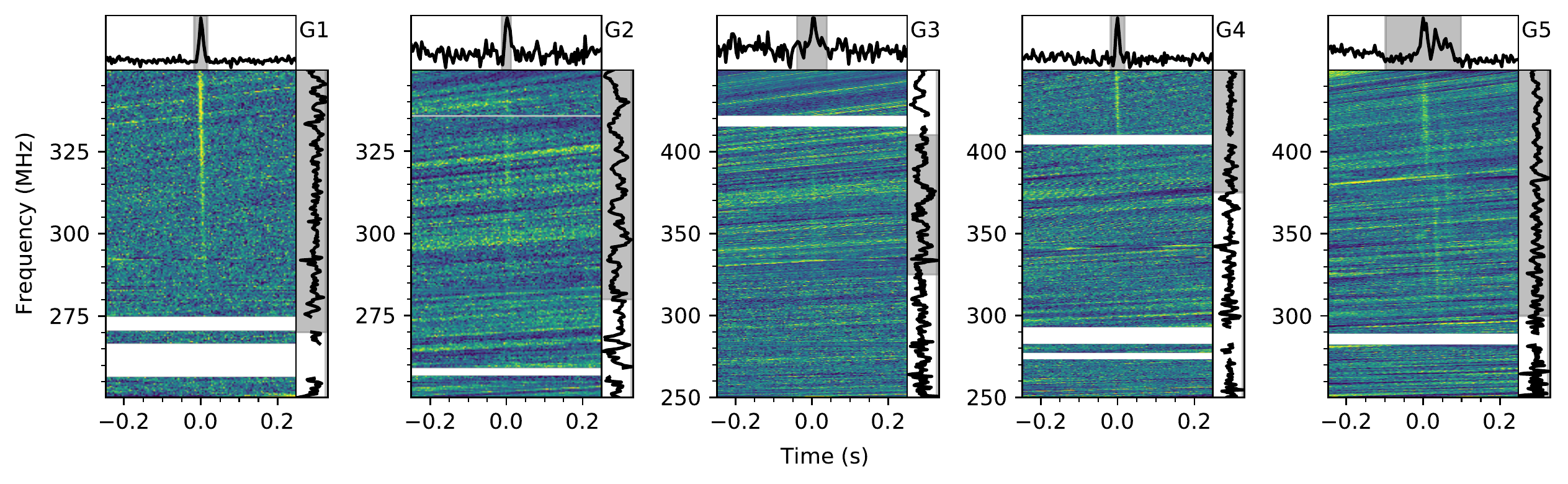}
    \caption{Dynamic spectra of bursts detected with the uGMRT. These dynamic spectra have been averaged to a time resolution of 3.93\,ms and a frequency resolution of 0.391\,MHz. Note the different receiver bandwidth for bursts G1--2 and G3--5. 
    Otherwise, the plot features are the same as in Figure~\ref{fig:lofar_waterfalls}.  The diagonal striations in the bandpass are due to residual broad-band RFI.}
  \label{fig:gmrt_waterfalls}
\end{figure*}

We detected five bursts in the uGMRT data. Two in the 250--350-MHz band on 2020\,June\,15 and three in the 250--450-MHz band on 2020\,July\,1. We did not detect any bursts on 2020\,May\,29. The burst dynamic spectra are presented in Figure~\ref{fig:gmrt_waterfalls}. To calibrate the bursts we have used the counts per Jansky estimated for every clean frequency channel from the ``ON'' and ``OFF'' data of 3C48. The conversion factors are multiplied with the filterbank counts to get the calibrated data. The measured peak flux densities and fluences are presented in Table~\ref{tab:lofar_burst_properties}.  

\begin{figure}
    \centering
    \includegraphics{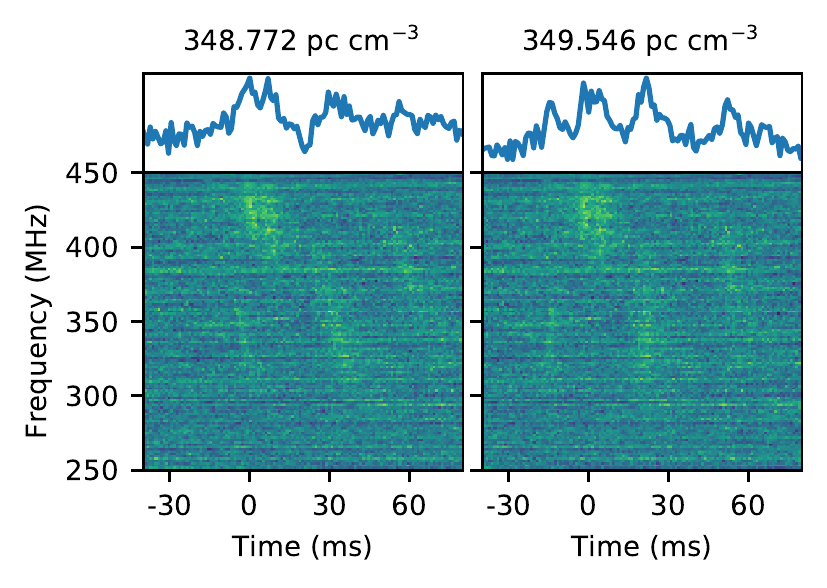}
    \caption{Burst G5 dedispersed to the best-fit DM from \citet{nimmo2020} (left) and the optimum DM from aligning the sub-bursts (right; see main text).} 
    \label{fig:g5_dmopt}
\end{figure}

We caution the reader not to over-interpret the spectral structure of the uGMRT bursts, due to the baseline variations from residual broad-band RFI in the dynamic spectra. The four sub-bursts in burst G5, however, are likely of astrophysical origin: the dispersion delay (quadratic and $\sim$16~s from 450 to 250~MHz) and dispersion smearing within the frequency channels of these candidates are as expected for the DM of the source and we find no candidate bursts at other DMs with similar properties. Using \texttt{DM\_phase}\footnote{\url{https://github.com/danielemichilli/DM_phase}} \citep{dmphase} we align the sub-bursts of G5 and find an optimum $\mathrm{DM}=349.5\pm0.1$ pc cm$^{-3}$. In Figure~\ref{fig:g5_dmopt} we show a comparison of the dynamic spectra of the bursts, dedispersed to the fiducial and optimum DM values.

\subsection{CHIME/FRB}\label{ssec:results_chime}

\begin{figure*}
  \centering
    \includegraphics[width=0.95\textwidth]{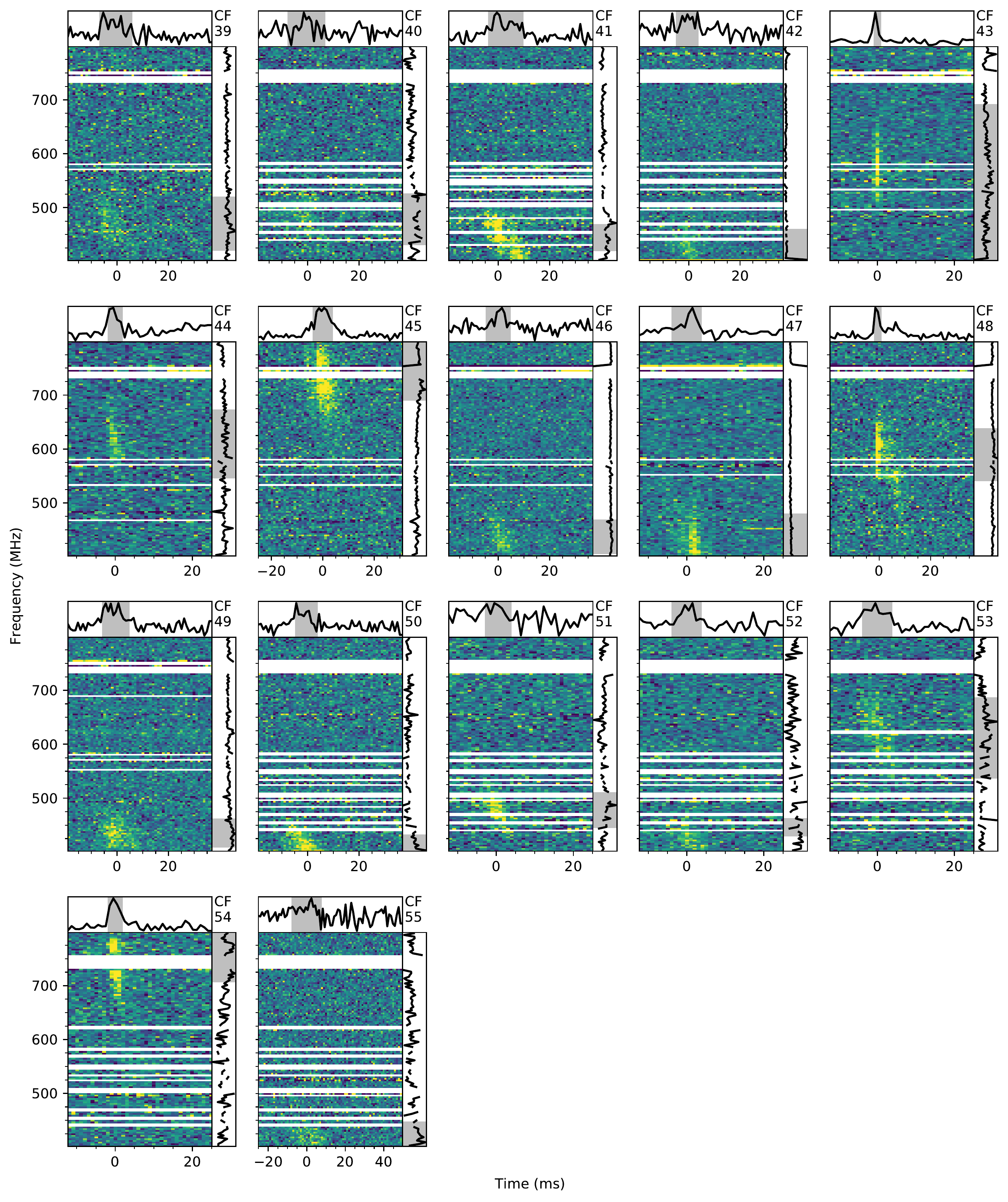}
    \caption{Dynamic spectra of bursts detected with CHIME/FRB. These dynamic spectra have time resolution
    of 0.98304\,ms and have been averaged to a frequency resolution of 3.125\,MHz. Otherwise, the plot features are the same as in Figure~\ref{fig:lofar_waterfalls}.}
  \label{fig:chimefrb_waterfalls}
\end{figure*}

We present 17 new bursts detected by CHIME/FRB since \periodicitypaper\footnote{Note that the burst arrival times were previously announced at \url{https://www.chime-frb.ca/repeaters/180916.J0158+65}.}. The burst dynamic spectra are presented in Figure~\ref{fig:chimefrb_waterfalls}. The morphologies of the bursts are comparable to those previously detected by CHIME/FRB; they exhibit narrow (50--150\,MHz) bandwidths and sometimes show downward-drifting sub-bursts \citep{chi19_8repeaters}.

We construct burst models and measure peak fluxes and fluences as in \citet{r3_period2020}. In summary, we fit single- or multi-component models of dynamic spectra to the 0.98304-ms
total-intensity data for each burst using a least-squares algorithm. Data are flux calibrated using transits of steady sources and scaled
by the beam response using the best known location of \rthree\ and a model for the synthesized beams. For bursts detected outside of the FWHM of the synthesized beams at 600\,MHz, peak fluxes and fluences are lower limits. 

\subsection{Frequency dependence of periodic activity}
\label{sec:periodic_activity}

We recalculate the activity period and burst rate of \rthree\ 
using the same methods as described in \periodicitypaper.
We measure the activity period to be $16.33\pm0.12$\,days,
with a 5.2-day window, based on the now reported 55
CHIME/FRB detections. We use reference MJD $\phi_0 = 58369.40$ to put the
average arrival time of the CHIME/FRB bursts at $\phi=0.5$. We estimate
the CHIME/FRB detection rate to be $0.8\pm0.3$ bursts
per hour above a fluence threshold of 5.1 Jy ms for a $\pm2.6$-day
interval around each cycle of activity (31 detections in 39.1 hours of
exposure). In the $1\sigma$ activity window of $\pm$0.96 days around
each cycle of activity, we estimate the detection rate to be
$1.5^{+0.8}_{-0.6}$ bursts per hour (22 detections in 14.9 hours of
exposure). 

\begin{figure}
  \centering
   \includegraphics[width=\columnwidth]{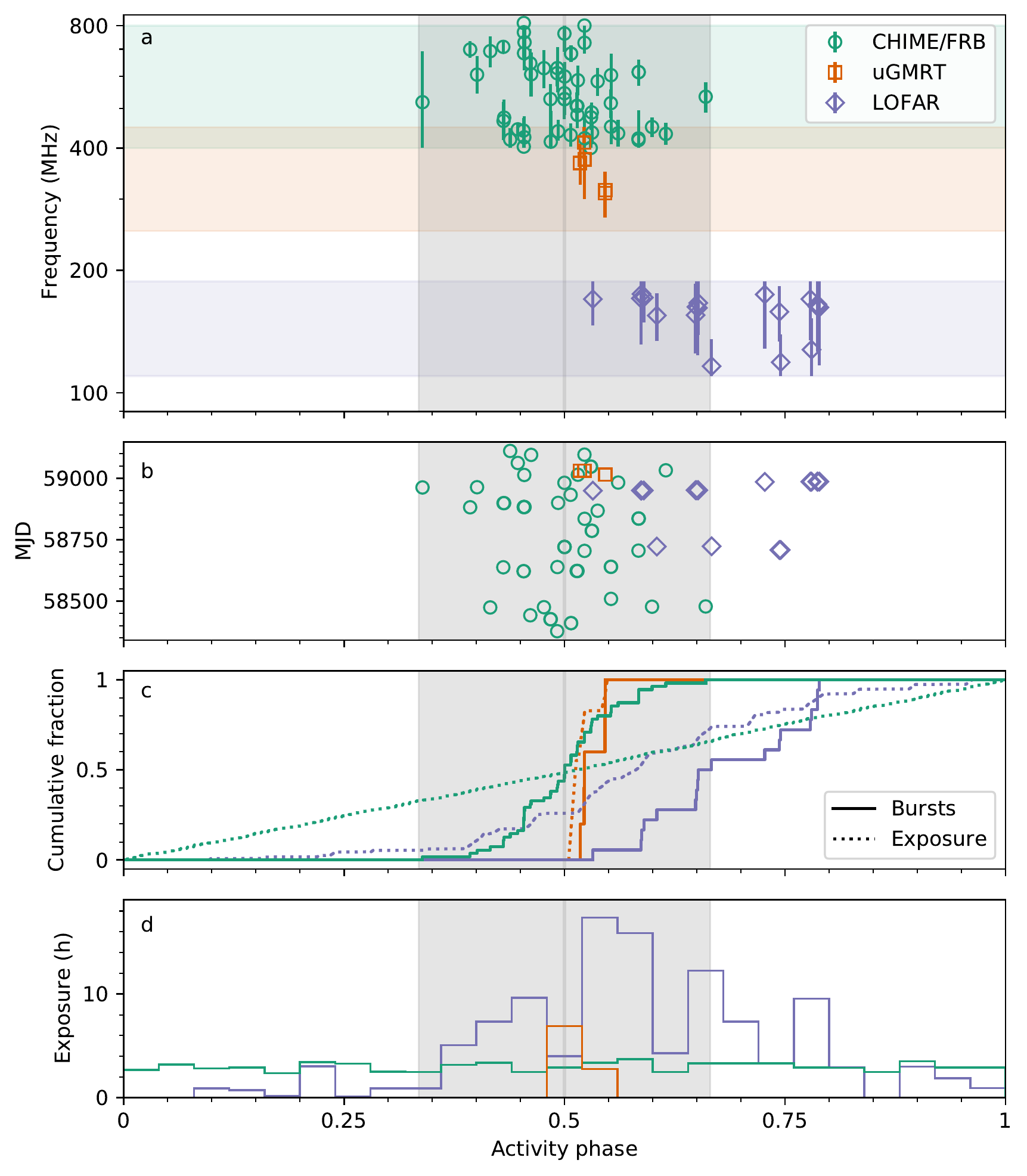}
    \caption{The activity phases of CHIME/FRB (green circles), uGMRT (orange squares) and LOFAR (purple diamonds) bursts folded
      on the 16.33-day activity period of \rthree. Panel (a) shows
      the activity phase of the bursts versus observing frequency. For each
      burst the spectral width is indicated by the error bars. Panel (b) shows 
      the burst MJDs versus activity phase. The cumulative 
      fraction of the number of bursts and the exposure are shown against 
      activity phase in Panel (c), while Panel (d) displays the exposure as a histogram. 
      The color coding is identical in the three panels. Whereas the CHIME/FRB 
      exposure is almost uniform with activity phase, the exposure of the LOFAR
      observations is focused predominantly on the CHIME/FRB activity
      window. As the number of uGMRT observations is limited, the phase
      of the bursts is dominated by the phase of the observations.}
  \label{fig:burst_phase}
\end{figure}

Both LOFAR and uGMRT predominantly observed \rthree\ during the 5.4-day
activity window from \periodicitypaper, though in later activity
cycles LOFAR also targeted activity phases outside of the activity
window (see Figure~\ref{fig:obs}). Figure~\ref{fig:burst_phase} shows
the barycentred arrival times of the CHIME/FRB, uGMRT and LOFAR bursts
folded on the activity period of $16.33$\,days --- as a function of
observing frequency, MJD, as well as cumulative fractions and histograms of
the exposure of each instrument. The bursts observed with LOFAR fall
between activity phases of $0.53<\phi<0.79$, corresponding to a range
of 4.1\,days within the 16.33-day activity period. This observed LOFAR
activity window is nominally shorter than the 5.2-day activity window width
observed from the CHIME/FRB bursts, but additional LOFAR observations may show the activity window to be wider. The average activity phase of the
LOFAR bursts is $\phi\sim0.66$, which is offset by 2.6\,days from the
average activity phase of \rthree\ determined from CHIME/FRB bursts
(\periodicitypaper). We note that the LOFAR burst detections as a function of activity phase are not just simply a reflection of the observing exposure. Through a two-sample Kolmogorov-Smirnov test we can reject the null hypothesis that the cumulative distribution functions of the LOFAR exposure and the LOFAR bursts shown in Figure\,\ref{fig:burst_phase} are drawn from the same distribution ($p\sim10^{-10}$).

\begin{figure}
  \centering
   \includegraphics[width=\columnwidth]{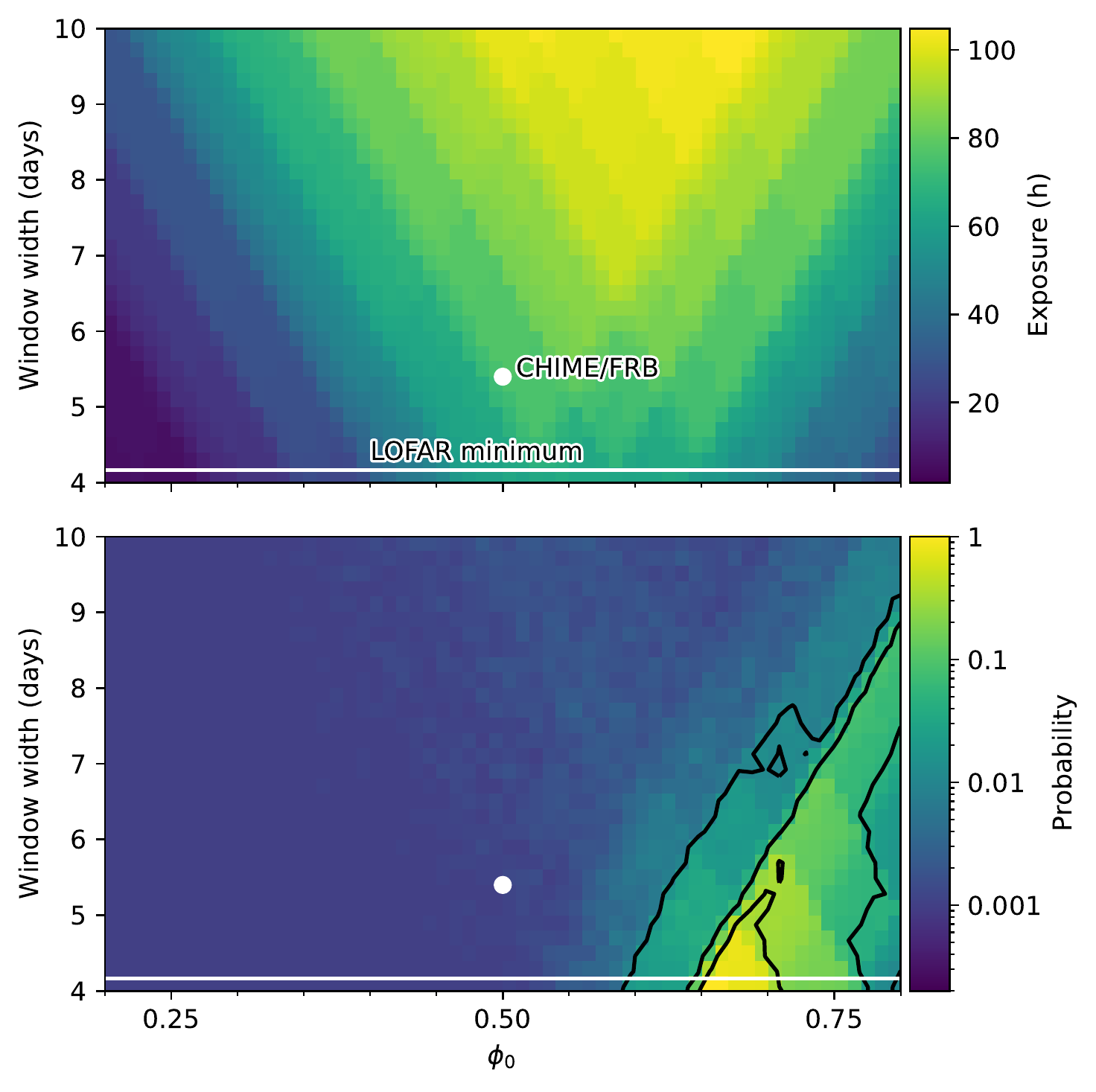}
    \caption{Constraints on the width $w$ and central phase $\phi_0$
      of the LOFAR activity window. The top panel shows the exposure
      covered by the LOFAR observations as a function of the activity
      window properties. The dot indicates the CHIME/FRB activity
      window properties, while the horizontal line denotes the minimum
      LOFAR activity window width set by the observed bursts. The
      bottom panel shows the constraints on the width and central
      phase of the LOFAR activity window, based on numerical
      simulations where burst times of arrival are drawn from a
      uniform distribution within the activity window and 18 bursts
      are coincident with the LOFAR observations and the observed
      activity phases ($0.53<\phi<0.79$). The contours provide the
      68\%, 95\% and 99\% confidence regions.}
  \label{fig:exposure_map}
\end{figure}

We performed numerical simulations to investigate the impact of the
non-uniform exposure of the LOFAR observations as a function of
activity phase (see Figure~\ref{fig:burst_phase}c and d) on the properties of
the activity window at LOFAR frequencies. We define an activity window
by its width $w$ and central phase $\phi_0$, and compute the effective
exposure of the LOFAR observations that fall within this activity
window. The top panel of Figure~\ref{fig:exposure_map} shows the LOFAR
exposure given the properties of the activity window. As 18 bursts
were observed with LOFAR, the exposure will provide the burst rate $r$
as a function of $w$ and $\phi_0$. Given the activity window
properties and the burst rate, we draw burst arrival times from a
uniform distribution within the activity window for the activity
cycles spanning the LOFAR observations (see Figure~\ref{fig:obs}). For
each $w$ and $\phi_0$ combination, we run mulitple simulations to
obtain the fraction of simulations where the simulated bursts fall
within both the LOFAR observations as well as the observed phase range
of LOFAR bursts ($0.53<\phi<0.79$). The bottom panel of
Figure~\ref{fig:exposure_map} shows this fraction, which we treat as
the probability that all simulated bursts fall within the observed
LOFAR activity phase range.

The simulations show that the observed activity window of the LOFAR
bursts is delayed with respect to the activity window observed by
CHIME/FRB. The best fit parameters for $w$ and $\phi_0$ are determined
through a Markov-Chain Monte Carlo analysis using the \texttt{emcee}
software \citep{fhd+13} to sample and maximize the probability that
all simulated bursts coincident with the LOFAR observations fall
within the observed LOFAR activity range. Flat priors were used for
both parameters, though the width was limited to the observed minimum
range of $w>4.1$\,days. The posterior distributions were obtained
using 32 walkers for 20000 steps, well beyond 100 times the largest
autocorrelation of the fitted parameters. After discarding a burn-in
phase of 1000 steps, and thinning by 35 steps, we obtained
$\phi_0=0.72_{-0.04}^{+0.07}$ and $w=5.0_{-0.8}^{+2.3}$\,d. These
values correspond to a LOFAR burst rate of
$r=0.32_{-0.04}^{+0.08}$\,hour$^{-1}$ for a fluence limit of 26\,Jy\,ms.

\section{Discussion} \label{sec:discussion}

\subsection{Lowest-frequency emission}

\rthree\ already held the record for the FRB with the lowest-frequency emission detected to date. Using the GBT and SRT, respectively, \citet{chawla2020} and \citet{pilia2020} previously presented a total of 10 burst detections at frequencies as low as 300\,MHz. Here we present 18 bursts detected in the 110--188-MHz band.

Notably, the majority of these bursts are brighter in the top-half of the band, but burst CV4 clearly demonstrates that emission can be detected down to at least 110\,MHz (Figure~\ref{fig:lofar_waterfalls}).  Searches for \rthree\ emission at radio frequencies $< 100$\,MHz are thus well motivated, and are underway using joint observations with LOFAR and NenuFAR \citep[New Extension in Nan\c{c}ay Upgrading LOFAR;][]{bondonneau2020}.  These observations can better quantify whether the observed burst rate is systematically reduced at lower radio frequencies.  While only three of the 18 bursts we have observed here are emitting predominantly in the lowest part of the LOFAR HBA band ($110-140$\,MHz), one must take observational biases like increased sky temperature ($T_{\rm sky}$) and larger pulse width into account (our detection metric scales as $F/\sqrt{w}$, where $F$ and $w$ are burst fluence and width, respectively).  The current sample is too small to make robust statements about the burst rate declining towards the bottom of the LOFAR HBA band.

Similarly, the detection of \rthree\ at $110-188$\,MHz provides renewed hope for detections in wide-field, low-frequency surveys \citep[e.g.,][]{scb+19}.

The simple fact that \rthree\ is visible at 110\,MHz sets new requirements on the emission mechanism and constraints on the effect of free-free absorption in the local medium.  Based on the lack of free-free absorption at 300\,MHz, \citet{chawla2020} argued that \rthree\ is not associated with a hyper-compact H\textsc{ii} region or a young ($< 50$\,yr) supernova remnant. This is consistent with the lack of local H$\alpha$ emission or a persistent radio counterpart, as shown by \citet{tendulkar2020} and \citet{marcote2020}, respectively.  Here we show that the circumburst environment is optically thin to free-free absorption at 110\,MHz.  Considering an ionized nebula of size L$_{\text{pc}}$, and  DM $<$ 70 pc cm$^{-3}$ \citep{marcote2020}, we use the following expression for the optical depth due to free-free absorption \citep{condon2016essential}:
\begin{equation}
\begin{aligned}
\tau_{\text{ff}} = 1.6 \times 10^{-3} \times \Big( \frac{\textrm{T}}{10^{4} \text{K}} \Big)^{-1.35} \times \Big( \frac{\nu}{1 \text{GHz}} \Big)^{-2.1} \\
\times \frac{1}{\text{f}_{\mathrm{eff}} \mathrm{L}_{\text{pc}}} \times \Big( \frac{\mathrm{DM}}{70 \text{ pc cm}^{-3}}\Big)^{2} \ll 1,
\label{Eq:ff}
\end{aligned}
\end{equation} 
where $\text{f}_{\mathrm{eff}}$ accounts for the volume-filling factor and the electron density fluctuation in the circum-burst medium. For $\text{f}_{\mathrm{eff}} = 1$, we find L $\gg$ 0.16 pc (T/$10^4$~K)$^{-1.35}$. The Crab nebula, for comparison, is about 1.7\,pc in radius.
\citet{piro2016} discusses how a surrounding supernova remnant can absorb bursts at low radio frequencies.  In the context of that work (see their Figure~4), our detection of a burst down to 110\,MHz implies a source age of at least $100-600$\,yrs, for assumed ejecta masses between $3-10\,\msun$. Similarly, \citet{marcote2020} discuss how the relatively low RM and lack of a persistent radio counterpart constrain the age of \rthree\ to be $\geq 300$\,yrs, in the context of models that describe it as a young flaring magnetar in a dense nebula \citep{metzger2019}.

Models that describe FRBs via synchrotron maser emission from decelerating relativistic blast waves predict that the intrinsic FRB fluence is lower at lower radio frequencies \citep{metzger2019}.  In contrast, at LOFAR frequencies of $110-188$\,MHz we see $>10\times$ higher fluences compared to bursts detected by CHIME/FRB at $400-800$\,MHz (Table~\ref{tab:lofar_burst_properties}).  Similarly, the detection of 110-MHz emission also places constraints on magnetospheric emission heights and models, e.g., the curvature radiation model \citep{kumar2017}, where a particle density $\sim10^{17}$ cm$^{-3}$ is necessary to produce a typical FRB luminosity ($L_\mathrm{iso} \sim 10^{43}$ erg s$^{-1}$) whereas the particle density can also not be much larger than this value for the emitting region to remain transparent to $\sim$100-MHz emission.

\subsection{Spectro-temporal and polarimetric behaviour}

The spectro-temporal properties of the 17 CHIME/FRB bursts presented here (Figure~\ref{fig:chimefrb_waterfalls}) are consistent with those of the 38 CHIME/FRB bursts previously presented in the literature.  As such, we do not discuss their properties in detail.

The five new uGMRT bursts (detected at $200-450$\,MHz) complement the 15 previously detected at $550-750$\,MHz \citep{marthi2020}.  Of the new uGMRT bursts, G5 (Figures~\ref{fig:gmrt_waterfalls} and \ref{fig:g5_dmopt}) has the most intriguing spectro-temporal properties, showing sub-structures that do not completely match the typical `sad trombone' downward-drift features seen from repeaters \citep[e.g.,][]{hessels19}.  A similar morphology has been observed in some bursts from \rone\ \citep{caleb2020,hilmarsson2020} and \rthree\ \citep{chawla2020}.  However, it is unclear whether these represent drifting sub-bursts within a single burst envelope, or a closely spaced set of separate bursts peaking at different frequencies, and then each individually drifting downwards in frequency.  The short-timescale sub-structure of burst G5 also allows us to better investigate possible DM variations (as we discuss in the next sub-section).

The 18 LOFAR bursts constitute our first robust view of FRB emission below 300\,MHz. These bursts have large widths (40 to 160\,ms; Table~\ref{tab:lofar_burst_properties}) compared to other FRBs in general \citep[e.g., Figure~13 in][]{petroff2019}, and compared to \rthree\ at high frequency \citep{nimmo2020}. This broadening is not due to dispersion smearing (see~\S\ref{sec:obs:lofar} and the coherently dedispersed waterfalls in Figure~\ref{fig:fitburst}). For comparison, the typical burst width at 1.7\,GHz is $\sim2-3$\,ms, and burst sub-structure has been detected on a wide range of timescales, down to $\sim3-4$\,$\upmu$s \citep{nimmo2020}. It is not completely clear whether the larger burst widths at lower frequencies are simply due to scatter broadening (extrinsic multi-path propagation through the intervening material), or whether they reflect an intrinsic aspect of the emission process --- like the `sad trombone' effect seen from \rthree\ and other repeaters, where the sub-burst drift to later times increases towards lower frequencies \citep{hessels19}.  

Previously, \citet{marcote2020} estimated a scattering time of 2.7\,$\upmu$s at 1.7\,GHz, and \citet{chawla2020} constrained the scattering time to be $<1.7$\,ms at 350\,MHz (this is compatible with the higher-frequency measurement).  For a scattering time that scales with frequency as $\tau_{\rm scatt} \propto \nu^{-4}$, the effect is $30 \times$ larger at 150\,MHz compared to 350\,MHz and $100 \times$ larger at 110\,MHz.  Based on the measured scattering time of \citet{marcote2020}, we expect a scattering time of $\sim40$\,ms at 150\,MHz, for a $\nu^{-4}$ scaling.  This matches well with the modeled scattering times of the high-S/N complex voltage data bursts CV1 and CV4 (Table~\ref{tab:cv_parameters}).  Regardless of modeling, bursts I14 and CV4 are the brightest detected bursts in the top- and bottom-half of the LOFAR HBA band, respectively, and both show clear asymmetric tails (as do other high-S/N bursts).  Their measured widths of $54\pm2$\,ms at 161\,MHz (I14) and $158\pm7$\,ms at 116\,MHz (CV4) are also consistent with what one would predict by extrapolating from the scattering measurement of \citet{marcote2020}. However, the LOFAR bursts may be additionally broadened by a poorly resolved `sad trombone' effect as well.  The drift towards later times increases towards lower frequencies \citep{hessels19,josephy2019} and, based on \rthree\ observations in the CHIME/FRB band (PR3), could be $\sim10$\,ms per $\sim50$\,MHz at LOFAR frequencies. We also note that many bursts from \rone\ show asymmetric burst profiles regardless of scattering \citep{hessels19}. Scatter-broadening can blend multiple intrinsic burst components \citep[see also \S4.3 of][]{day2020}. Additional LOFAR observations spanning a wide range of epochs can constrain whether the scattering time varies, as is seen from the Crab pulsar \citep{driessen2019}.  Our burst detections spanning $\sim10$ months show no obvious evidence for this, however.

The detection of \rthree, with DM $\sim350$\,pc~cm$^{-3}$, in the LOFAR HBA band contrasts with results from the LOFAR HBA ($110-188$\,MHz) census of slow and millisecond pulsars \citep{bkk+16,kvh+16}, as well as the new and known pulsars detected in the LOFAR Tied-Array All-Sky survey (LOTAAS; \citealt{scb+19}).  None of these observations detect Galactic pulsars with $\mathrm{DM}\gtrsim220$\,pc\,cm$^{-3}$. The absence of LOFAR pulsar detections above this limit is consistent \citep{scb+19} with predictions from the Galactic scattering relations derived by \citet{bcc+04} and \citet{gkk+17}.  Indeed, extragalactic FRBs are often significantly less scattered than Galactic pulsars of comparable DM \citep[see, e.g., Figure~16 of][though note that \rthree\ has low Galactic latitude $b = 3.7^\circ$, towards Galactic longitude
$l=129.7\degr$]{cordes2019}.

In previous, simultaneous LOFAR HBA, GBT and CHIME/FRB observations, bursts were detected at $>300$\,MHz, but not at $110-188$\,MHz \citep{chawla2020}. Of the CHIME/FRB, uGMRT and LOFAR observations presented here, the arrival times of five CHIME/FRB bursts from \citet{r3_period2020} overlap with LOFAR observations, of which three have been presented in \citet{chawla2020}. None of these bursts have counterparts in the $110-188$\,MHz band of LOFAR. Similarly, \citet{pearlman2020} used simultaneous CHIME/FRB and Deep Space Network (DSN) 70-m dish observations to demonstrate a burst detection in the CHIME/FRB band, but none at 2.3\,GHz or 8.4\,GHz. Clearly, \rthree\ bursts have a low instantaneous bandwidth, as has been seen for \rone\ \citep{gourdji2019,majid2020} and beautifully demonstrated for FRB~20190711A \citep{kumar2021}. The LOFAR HBA bursts we present here show bandwidths of $20-50$\,MHz, and are consistent with an emerging picture in which repeating FRB bursts have typical fractional bandwidths ($\mathrm{BW}/\nu_{\rm obs}$) of $\sim20$\% \citep[e.g.,][]{chi19_8repeaters,gourdji2019}.  The large sample of \rthree\ bursts that are now available from $110-1700$\,MHz could also be stacked to determine an average spectral index --- though that requires careful consideration of CHIME/FRB beam effects and other instrumental biases/selection effects, and is beyond the scope of this work.  

Lastly, for three LOFAR HBA bursts with full polarimetric data available, we have measured the linear and circular polarization fractions (Figure~\ref{fig:pol_profiles}).  We find broad similarity to the polarimetric properties measured for \rthree\ bursts at 350\,MHz \citep{chawla2020} and 1.7\,GHz \citep{nimmo2020}: i.e., the high linear polarization fraction, with flat polarization angle during the burst, and negligible circular polarization fraction, persists from $130-1700$\,MHz --- almost four octaves in radio frequency.  For the low-frequency Burst CV4, the lower linear polarization fraction is likely due to scattering, though investigations are ongoing to determine whether this could be the result of Faraday conversion \citep{gruzinov2019,vedantham2019}.  The remarkably similar polarimetric burst profiles across a wide range of radio frequencies is dissimilar to what is seen in some pulsars \citep{noutsos2015}, and provides a novel constraint on FRB emission theory.

\subsection{DM and RM variations}

Despite the low frequencies of our LOFAR HBA burst detections, the large burst widths and likely presence of scattering make it difficult to precisely and accurately determine the burst DMs in order to investigate potential DM variability.  For this reason, we have dedispersed all bursts in Figure~\ref{fig:lofar_waterfalls} to a best-fit literature value of DM$=348.772$\,pc~cm$^{-3}$ \citep{nimmo2020}. Previously, \periodicitypaper\ searched for DM variability, also as a function of \rthree's activity phase.  Using structure-optimized DMs \citep{hessels19} from four bursts detected using the CHIME/FRB baseband capture system, they found no DM variations with magnitude $\gtrsim0.1$\,pc~cm$^{-3}$.

Using three high-S/N bursts detected at $550-750$\,MHz with uGMRT, and searching for a DM that maximizes burst sub-structure \citep{hessels19}, \citet{marthi2020} also found no evidence for large DM variations.  Their strongest constraint, DM $=348.8\pm0.1$\,pc~cm$^{-3}$ (for their burst 11), is also consistent with the DM$=348.772$\,pc~cm$^{-3}$ \citep{nimmo2020} value we use throughout this paper.  Optimizing the sub-structure of burst G5 (Figure~\ref{fig:g5_dmopt}), we find $\mathrm{DM}=349.5(1)$ pc cm$^{-3}$. However, as we noted in the previous sub-section, there is some ambiguity in the interpretation of the burst structure of G5. Nevertheless, any DM variations are constrained to be $\lesssim 1$ pc cm$^{-3}$. 

The LOFAR HBA burst detections do provide precise RM values. Comparing with previous GBT and CHIME/FRB measurements \citep{chi19_8repeaters,chawla2020}, and correcting for the variable RM contribution from the Earth's ionosphere, we find that \rthree\ shows significant variations at the level of $\sim2-3$\,rad~m$^{-2}$ ($\sim2$\% fractional; though the variable component of the RM may be much lower than the total line-of-sight RM, in which case the fractional variations could be much larger).  With only seven measurements in hand, it is not yet clear whether the observed RM is correlated with the activity phase of the source, or varies stochastically (Figure~\ref{fig:RM}).  Alternatively, the RM (and DM) could also depend on radio frequency, e.g., if there is a frequency dependence to the location of the emission region. Some pulsars are known to show variable apparent RM across their pulse profiles that indicates emission from multiple locations within the magnetosphere \citep[e.g.,][]{dai2015,ilie2019}. Similar effects might be seen in FRBs if their emission is magnetospheric in origin as well.

\subsection{Observed activity}

A revised analysis of the 38 previously published and 17 newly presented CHIME/FRB bursts --- spanning two years, and thus 45 activity cycles, from 2018 September 16 until 2020 September 19 --- provides a refined source activity period of $P_{\rm act} = 16.33 \pm 0.12$\,days using the same methodology as in \periodicitypaper.  This is consistent with the previously published determination of $P_{\rm act} = 16.35 \pm 0.15$\,days. The refined activity window is 5.2 days, as compared to the 5.4-day window that was measured using only 38 events.

Folding all bursts modulo $P_{\rm act}$, we see that the 18 LOFAR HBA bursts are systematically delayed in activity phase, by $\sim$three days (0.2 cycles), compared to the 55 CHIME/FRB bursts (Figure~\ref{fig:burst_phase}).  As we show in Figure~\ref{fig:exposure_map}, this effect is not simply a reflection of the observational exposure.  LOFAR HBA bursts have been detected in four activity cycles, namely Cycles 20, 21, 35 and 37.  Though the number of observations and bursts per cycle is low, these are consistent with the delayed activity being a time-invariant effect.  Cycle 37 provides the best single-cycle observational coverage and burst sample between CHIME/FRB and LOFAR (Figure~\ref{fig:obs}), and is also consistent with the overall picture one obtains by summing over all activity cycles.  An ongoing campaign of NenuFAR, LOFAR HBA and Effelsberg observations will better characterize this frequency-dependent activity, and determine whether the low-frequency activity window has a larger duty cycle, or not.  A larger burst sample can also, in principle, determine the functional form of the activity delay with frequency, e.g., whether the delay as a function of frequency is linear or quadratic.

Higher-frequency detections of \rthree\ at L-band seem to preferentially arrive at the start of the CHIME/FRB-derived activity window (PR3, \citealt{aggarwal2020}). In follow-up observations of repeaters with periodic activity it thus seems wise to cover a broad window and not only the peak of activity derived at a different frequency.

\subsection{A self-consistent model for \rthree}

We now consider what models can naturally accommodate the wealth of observational facts available for \rthree.

Recently, \citet{tendulkar2020} used \emph{Hubble Space Telescope} and Gran Telescopio Canarias observations to demonstrate that \rthree\ is significantly offset with respect to nearby star-forming regions in its host galaxy.  Assuming it is a neutron star formed in one of these regions, as opposed to \emph{in situ}, it is more likely to be an old (100\,kyr to 10\,Myr) source.  Coupled with the observed 16.3-day activity period, this led \citet{tendulkar2020} to suggest that the system could be a high-mass X-ray binary (HMXB), where the bursts are possibly generated through an interaction between the companion wind and neutron star magnetosphere \citep[sometimes called a ``cosmic comb'' model\footnote{In this putative process, the ram pressure of the companion wind would in places exceed the magnetic pressure of the magnetospheric field lines and create a sheath of plasma wherein magnetic reconnections are triggered that accelerate particles to relativistic speeds. These particles could produce the coherent radio emission.}, e.g.,][]{zhang2017,iokazhang2020}. Furthermore, the detection of a wide range (factor $\sim1000$) of emission timescales and microsecond structure in some \rthree\ bursts \citep{nimmo2020} is more naturally explained in terms of magnetospheric emission models, as opposed to those that invoke emission in a relativistic shock far from the neutron star.  

HMXBs are relatively common: there are roughly 200 known in the Milky Way \citep{coleiro2013,walter2015}.  Furthermore, studies find that binary interaction dominates the evolution of massive stars \citep{sana2012}, and hence highly magnetized neutron stars in such binaries should not be particularly rare.  In fact, HMXBs are arguably {\it too} abundant to explain repeating FRBs, unless one invokes particular evolutionary stages, sporadic emission episodes, and/or viewing geometry to explain the lack of radio burst detections from Galactic HMXBs, as we discuss further below.  

The detection of low-frequency bursts with no simultaneous high-frequency emission \citep{pearlman2020}, strongly challenges models
in which
the 16.3-day observed periodicity is the result of absorption by the companion wind \citep{lbg2020}.  Rather, our LOFAR HBA detections and the lack of any observed DM variations,
with $\Delta$DM~$\lapp 0.1$~pc~cm$^{-3}$ throughout the active window (\periodicitypaper), 
suggest that we have a relatively clean line-of-sight to the burst source itself.  Nonetheless, the observed RM variations of \rthree\ are atypical and not seen from isolated Galactic pulsars; they suggest that it is in a special local environment.

It is interesting to compare \rthree\ with known radio pulsar/OB-star binary systems.  One example is the Small Magellanic Cloud (SMC) radio pulsar/B1V-companion binary system PSR~J0045$-$7319. 
In this 51-day highly eccentric
binary, at periastron and apastron, the pulsar approaches to within four and 34 B-star radii from the companion, respectively, probing very different regions of the stellar wind.  Yet
\citet{ktm96} found an upper limit on
DM variations $<0.9$\,pc~cm$^{-3}$, yielding a strong constraint on the stellar wind, $<10^{-11}$~M$_{\odot}$~yr$^{-1}$.
Given the low metallicity of the SMC\footnote{We note that the \rthree\ host galaxy has a similar oxygen abundance as that of \ion{H}{2} regions in the SMC \citep{tde+17,tendulkar2020}}, a B-star with so weak a wind is not unexpected.  

In a similar way, the lack of large observed DM variations (i.e., $\Delta\mathrm{DM}\lesssim1$\,pc~cm$^{-3}$) from \rthree\ may not be problematic for a massive star binary model.  Based on population synthesis arguments, \citet{zhanggao2020} argue that B-type stars would be the most likely companions to FRB binary sources.  If the putative \rthree\ orbit is not very eccentric, the variation in pulsar/companion distance with orbital phase may not be large.  This, together with a relatively weak companion wind, may make DM variations hard to observe.
On the other hand,
the Galactic eccentric radio pulsar/OB-star binary
PSR J1740$-$3052 does show DM variations of order $\sim$2~pc~cm$^{-3}$ near periastron in its 231-day orbit; 
these imply a mass-loss rate on the order
of $10^{-9}$~M$_{\odot}$~yr$^{-1}$, still low for a Galactic O or early B star \citep{msk+12}.  For similar strength wind, given the 16.3-day orbital period and hence closer pulsar/companion distance, the CHIME/FRB-active window would have to occur near apastron, or the orbit would have to be fairly circular to avoid detectable DM changes.
Similarly, the Galactic highly eccentric 1237-day
radio pulsar/Be-star binary PSR~B1259$-$63
shows large DM, RM and scattering-time variations in the $\sim$50 days near periastron ($\Delta\mathrm{DM}\simeq 6-8$\,pc~cm$^{-3}$, $\Delta\mathrm{RM}\simeq6000$\,rad~m$^{-2}$, and increased scattering such that 1.5-GHz pulse profiles become unobservable for this 48-ms pulsar),
though these are likely due to the presence of a circumstellar disk through which the pulsar passes \citep{jml+92,jml+96,jbwm05}. An emission-line star in the \rthree\ system seems implausible, because of the shorter orbital period, hence more compact orbit size and pulsar/star separation, together with the constraints on DM variations, absence of sizable RM variations, and given our new low-frequency detections which rule out large scattering times and which are offset in phase. 

In the context of interacting binary models \citep{lbg2020,iokazhang2020,popov2020,du2021}, the FRBs may be produced by a highly magnetized neutron star whose magnetosphere is `combed' by the ionized wind of a massive companion star.  Such interaction could lead to magnetic reconnection events, which have been proposed as a source of FRBs \citep{lp2020}.  The bursts may only be visible within a funnel where the neutron star's wind shields against the companion's wind, which is otherwise opaque to induced Compton or Raman scattering for FRB emission \citep[][see their Figure~1]{iokazhang2020}.  Windows of observable burst activity, lasting for $\sim4-5$ days in the case of \rthree, then correspond to when this funnel, and the induced magnetic tail of the neutron star, are pointed towards Earth.  This special viewing geometry could also, in principle, explain why Galactic HMXBs are not known to be prolific sources of bright radio bursts.

The funnel and magnetic tail 
can also be swept back by orbital motion. \citet{wang2019} and \citet{lyu2020} discuss a radius-to-frequency mapping model to explain the `sad trombone' effect seen from repeating FRBs on timescales of milliseconds \citep{hessels19}.  If the radio emission frequency of bursts indeed scales with distance from the central source, where the magnetic field strength and plasma density are lower at larger distances, then this could plausibly also explain why the LOFAR HBA bursts we have observed are delayed in activity phase with respect to those seen at higher frequencies by CHIME/FRB.  In such a toy model, the LOFAR bursts would originate at larger distances from the neutron star, in a swept-back magnetic tail.

While a model in which \rthree\ is a highly magnetized neutron star in an interacting HMXB system can plausibly explain all the observed phenomena to date, several authors have argued that the 16.3-day activity period of \rthree\ is the rotational period \citep{beniamini2020}, or precession period \citep{levin2020,sobyanin2020,yangzou2020,zanazzi2020}, of an isolated magnetar.  \citet{tendulkar2020} have argued against a young source, and the consistent polarization position angle \emph{between} bursts \citep{nimmo2020} sets significant constraints on precession models.  Nonetheless, in the context of these non-binary models, the frequency dependence of the observed activity could also be interpreted as a radius-to-frequency mapping effect in an emission cone that is slowly sweeping past the line of sight and is perhaps swept backwards at higher altitudes.

Lastly, we note that \rthree\ and \rone, the two best-studied repeaters, share remarkably similar phenomenology (see \S\ref{sec:intro}) --- despite the fact that \rone\ is hosted in a much less massive dwarf galaxy and is coincident with a compact, persistent radio source.  They are almost certainly of a similar physical origin, though \rone\ may be in the vicinity of an accreting massive black hole \citep{michilli18}.  Such a Galactic-center-like environment, may explain why --- unlike \rthree\ --- \rone\ has only been detected once at radio frequencies $<1$\,GHz \citep{josephy2019}. Their different Galactic latitudes ($b=3.7^\circ$ for \rthree\ and $b=-0.2^\circ$ for \rone), might also play some role. Perhaps, in the context of an HMXB model, if the radio emission frequency is tied to the instantaneous altitude of the emission site, it is possible that the characteristic emission height is tied to the companion wind strength.

\section{Conclusions} \label{sec:conclusions}

Using LOFAR observations in the $110-188$-MHz band, we have detected 18 bursts from \rthree. Since some of the detected bursts are bright down to the lowest-observed frequency of 110\,MHz, it is likely that \rthree\ emission extends to even lower frequencies --- though scattering ($\sim50$\,ms at 150\,MHz) and sky background temperature will limit their detectability unless the fluences also increase and compensate for these effects.  We are now actively searching for $<100$\,MHz emission using coordinated LOFAR and NenuFAR observations. The discovery of FRB emission in the $110-188$-MHz band also gives new impetus to searches for additional sources in this band.

LOFAR polarimetric data demonstrate consistency with the properties previously presented at higher radio frequencies of $300-1700$\,MHz.  The LOFAR bursts, combined with previous measurements, also show $2-3$\,rad~m$^{-2}$ RM variations.  One highly structured burst from the five new detections we presented from uGMRT ($200-450$\,MHz) leaves room for small but significant DM variations ($\lesssim$1\,pc~cm$^{-3}$) depending on the interpretation of its burst morphology.

Lastly, we also presented 17 new CHIME/FRB bursts detected at $400-800$\,MHz.  For five CHIME/FRB bursts with overlapping LOFAR observations, we detect no emission in the LOFAR $110-188$-MHz band.  This again emphasises the narrow-band nature of repeating FRB bursts, as previously discussed for a number of sources in the literature.

Using the full available sample of 55 CHIME/FRB bursts, spanning two years, we confirm that \rthree\ is periodically active, with a refined period of $16.33\pm0.12$ days over the 45 cycles observed to date.  Comparing with the 55 CHIME/FRB bursts, we show that the LOFAR bursts arrive systematically later in the 16.33-day activity cycle of the source.  We find a $\sim$three day (0.2 cycle) shift across the two octaves in radio frequency from 600\,MHz to 150\,MHz.

We interpret these results in the context of the rich set of observational facts that are known about the \rthree\ burst properties and the local environment in its massive host galaxy.  We discuss how a model in which \rthree\ is an interacting neutron-star HMXB system can account for all the observational results to date. 

\acknowledgments{
We thank the anonymous referee for helpful comments that improved the quality of the manuscript.  We also thank Julian Donner and Caterina Tiburzi for their help with calibrating the LOFAR data.

This paper is based (in part) on data obtained with the International LOFAR Telescope (ILT) under project codes \texttt{LC12\_016}, \texttt{DDT12\_001}, \texttt{LC13\_016},  \texttt{DDT14\_005} and \texttt{COM\_ALERT}. LOFAR \citep{hwg+13} is the Low Frequency Array designed and constructed by ASTRON. It has observing, data processing, and data storage facilities in several countries, that are owned by various parties (each with their own funding sources), and that are collectively operated by the ILT foundation under a joint scientific policy. The ILT resources have benefitted from the following recent major funding sources: CNRS-INSU, Observatoire de Paris and Universit\'e d'Orl\'eans, France; BMBF, MIWF-NRW, MPG, Germany; Science Foundation Ireland (SFI), Department of Business, Enterprise and Innovation (DBEI), Ireland; NWO, The Netherlands; The Science and Technology Facilities Council, UK. 

We thank the staff of the GMRT that made the uGMRT observations possible. The GMRT is run by the National Centre for Radio Astrophysics of the Tata Institute of Fundamental Research.  

The CHIME/FRB Project is funded by a grant from the Canada Foundation for Innovation 2015 Innovation Fund (Project 33213), as well as by the Provinces of British Columbia and Qu\'{e}bec, and by the Dunlap Institute for Astronomy and Astrophysics at the University of Toronto. Additional support was provided by the Canadian Institute for Advanced Research (CIFAR), McGill University and the McGill Space Institute via the Trottier Family Foundation, and the University of British Columbia. The Dunlap Institute is funded by an endowment established by the David Dunlap family and the University of Toronto. Research at Perimeter Institute is supported by the Government of Canada through Industry Canada and by the Province of Ontario through the Ministry of Research \& Innovation. The National Radio Astronomy Observatory is a facility of the National Science Foundation operated under cooperative agreement by Associated Universities, Inc. We are grateful to the staff of the Dominion Radio Astrophysical Observatory, which is operated by the National Research Council Canada.

D.M. is a Banting Fellow.
J.W.T.H. acknowledges funding from an NWO Vici grant (``AstroFlash'').
P.C. is supported by an FRQNT Doctoral Research Award.
V.M.K. holds the Lorne Trottier Chair in Astrophysics \& Cosmology, a Distinguished James McGill Professorship and receives support from an NSERC Discovery Grant and Gerhard Herzberg Award, from an R. Howard Webster Foundation Fellowship from CIFAR, and from the FRQNT CRAQ.
M.B. is supported by an FRQNT Doctoral Research Award.
Y.G. acknowledges support from the Department of Atomic Energy, Government of India, under project \#12-R\&D-TFR-5.02-0700.
F.K. acknowledges support by the Swedish Research Council.
C.L. was supported by the U.S. Department of Defense (DoD) through the National Defense Science \& Engineering Graduate Fellowship (NDSEG) Program.
B.M. acknowledges support from the Spanish Ministerio de Econom\'ia y Competitividad (MINECO) under grant AYA2016-76012-C3-1-P and from the Spanish Ministerio de Ciencia e Innovaci\'on under grants PID2019-105510GB-C31 and CEX2019-000918-M of ICCUB (Unidad de Excelencia ``Mar\'ia de Maeztu'' 2020-2023).
P.S. is a Dunlap Fellow and an NSERC Postdoctoral Fellow.
K.S. acknowledges support by the NSF Graduate Research Fellowship Program.
FRB research at UBC is supported by an NSERC Discovery Grant and by the Canadian Institute for Advance Research. The CHIME/FRB baseband system is funded in part by a CFI John R. Evans Leaders Fund award to I.H.S.
}

\facilities{LOFAR, uGMRT, CHIME/FRB}

\software{\texttt{digifil} \citep{sb10}, \textsc{Presto} \citep{ran01}, \textsc{Dedisp} \citep{bbbf12}, \texttt{gptool} \citep{Susobhanan20}, \texttt{dspsr} \citep{sb10}, \texttt{ionFR} \citep{sot13}, \texttt{DM\_phase} \citep{dmphase}, \texttt{emcee} \citep{fhd+13}, \texttt{PSRCHIVE} \citep{psrchive}, \texttt{RM-Tools} \citep{rmtools}, Matplotlib \citep{matplotlib}, NumPy \citep{numpy}, Astropy \citep{astropy1, astropy2}, SciPy \citep{scipy}}

\bibliography{references}{}

\begin{thebibliography}{}
\expandafter\ifx\csname natexlab\endcsname\relax\def\natexlab#1{#1}\fi
\providecommand{\url}[1]{\href{#1}{#1}}
\providecommand{\dodoi}[1]{doi:~\href{http://doi.org/#1}{\nolinkurl{#1}}}
\providecommand{\doeprint}[1]{\href{http://ascl.net/#1}{\nolinkurl{http://ascl.net/#1}}}
\providecommand{\doarXiv}[1]{\href{https://arxiv.org/abs/#1}{\nolinkurl{https://arxiv.org/abs/#1}}}

\bibitem[{{Agarwal} {et~al.}(2020){Agarwal}, {Aggarwal}, {Burke-Spolaor},
  {Lorimer}, \& {Garver-Daniels}}]{agarwal2020}
{Agarwal}, D., {Aggarwal}, K., {Burke-Spolaor}, S., {Lorimer}, D.~R., \&
  {Garver-Daniels}, N. 2020, \mnras, 497, 1661, \dodoi{10.1093/mnras/staa1856}

\bibitem[{{Aggarwal} {et~al.}(2020){Aggarwal}, {Law}, {Burke-Spolaor}, {Bower},
  {Butler}, {Demorest}, {Linford}, \& {Lazio}}]{aggarwal2020}
{Aggarwal}, K., {Law}, C.~J., {Burke-Spolaor}, S., {et~al.} 2020, Research
  Notes of the American Astronomical Society, 4, 94,
  \dodoi{10.3847/2515-5172/ab9f33}

\bibitem[{{Astropy Collaboration} {et~al.}(2013){Astropy Collaboration},
  {Robitaille}, {Tollerud}, {Greenfield}, {Droettboom}, {Bray}, {Aldcroft},
  {Davis}, {Ginsburg}, {Price-Whelan}, {Kerzendorf}, {Conley}, {Crighton},
  {Barbary}, {Muna}, {Ferguson}, {Grollier}, {Parikh}, {Nair}, {Unther},
  {Deil}, {Woillez}, {Conseil}, {Kramer}, {Turner}, {Singer}, {Fox}, {Weaver},
  {Zabalza}, {Edwards}, {Azalee Bostroem}, {Burke}, {Casey}, {Crawford},
  {Dencheva}, {Ely}, {Jenness}, {Labrie}, {Lim}, {Pierfederici}, {Pontzen},
  {Ptak}, {Refsdal}, {Servillat}, \& {Streicher}}]{astropy1}
{Astropy Collaboration}, {Robitaille}, T.~P., {Tollerud}, E.~J., {et~al.} 2013,
  \aap, 558, A33, \dodoi{10.1051/0004-6361/201322068}

\bibitem[{{Astropy Collaboration} {et~al.}(2018){Astropy Collaboration},
  {Price-Whelan}, {Sip{\H{o}}cz}, {G{\"u}nther}, {Lim}, {Crawford}, {Conseil},
  {Shupe}, {Craig}, {Dencheva}, {Ginsburg}, {VanderPlas}, {Bradley},
  {P{\'e}rez-Su{\'a}rez}, {de Val-Borro}, {Aldcroft}, {Cruz}, {Robitaille},
  {Tollerud}, {Ardelean}, {Babej}, {Bach}, {Bachetti}, {Bakanov}, {Bamford},
  {Barentsen}, {Barmby}, {Baumbach}, {Berry}, {Biscani}, {Boquien}, {Bostroem},
  {Bouma}, {Brammer}, {Bray}, {Breytenbach}, {Buddelmeijer}, {Burke},
  {Calderone}, {Cano Rodr{\'\i}guez}, {Cara}, {Cardoso}, {Cheedella}, {Copin},
  {Corrales}, {Crichton}, {D'Avella}, {Deil}, {Depagne}, {Dietrich}, {Donath},
  {Droettboom}, {Earl}, {Erben}, {Fabbro}, {Ferreira}, {Finethy}, {Fox},
  {Garrison}, {Gibbons}, {Goldstein}, {Gommers}, {Greco}, {Greenfield},
  {Groener}, {Grollier}, {Hagen}, {Hirst}, {Homeier}, {Horton}, {Hosseinzadeh},
  {Hu}, {Hunkeler}, {Ivezi{\'c}}, {Jain}, {Jenness}, {Kanarek}, {Kendrew},
  {Kern}, {Kerzendorf}, {Khvalko}, {King}, {Kirkby}, {Kulkarni}, {Kumar},
  {Lee}, {Lenz}, {Littlefair}, {Ma}, {Macleod}, {Mastropietro}, {McCully},
  {Montagnac}, {Morris}, {Mueller}, {Mumford}, {Muna}, {Murphy}, {Nelson},
  {Nguyen}, {Ninan}, {N{\"o}the}, {Ogaz}, {Oh}, {Parejko}, {Parley}, {Pascual},
  {Patil}, {Patil}, {Plunkett}, {Prochaska}, {Rastogi}, {Reddy Janga},
  {Sabater}, {Sakurikar}, {Seifert}, {Sherbert}, {Sherwood-Taylor}, {Shih},
  {Sick}, {Silbiger}, {Singanamalla}, {Singer}, {Sladen}, {Sooley},
  {Sornarajah}, {Streicher}, {Teuben}, {Thomas}, {Tremblay}, {Turner},
  {Terr{\'o}n}, {van Kerkwijk}, {de la Vega}, {Watkins}, {Weaver}, {Whitmore},
  {Woillez}, {Zabalza}, \& {Astropy Contributors}}]{astropy2}
{Astropy Collaboration}, {Price-Whelan}, A.~M., {Sip{\H{o}}cz}, B.~M., {et~al.}
  2018, \aj, 156, 123, \dodoi{10.3847/1538-3881/aabc4f}

\bibitem[{{Barsdell} {et~al.}(2012){Barsdell}, {Bailes}, {Barnes}, \&
  {Fluke}}]{bbbf12}
{Barsdell}, B.~R., {Bailes}, M., {Barnes}, D.~G., \& {Fluke}, C.~J. 2012,
  \mnras, 422, 379, \dodoi{10.1111/j.1365-2966.2012.20622.x}

\bibitem[{{Bassa} {et~al.}(2017{\natexlab{a}}){Bassa}, {Pleunis}, \&
  {Hessels}}]{bassa2017a}
{Bassa}, C.~G., {Pleunis}, Z., \& {Hessels}, J.~W.~T. 2017{\natexlab{a}},
  Astronomy and Computing, 18, 40, \dodoi{10.1016/j.ascom.2017.01.004}

\bibitem[{{Bassa} {et~al.}(2017{\natexlab{b}}){Bassa}, {Pleunis}, {Hessels},
  {Ferrara}, {Breton}, {Gusinskaia}, {Kondratiev}, {Sanidas}, {Nieder},
  {Clark}, {Li}, {van Amesfoort}, {Burnett}, {Camilo}, {Michelson}, {Ransom},
  {Ray}, \& {Wood}}]{bassa2017b}
{Bassa}, C.~G., {Pleunis}, Z., {Hessels}, J.~W.~T., {et~al.}
  2017{\natexlab{b}}, \apjl, 846, L20, \dodoi{10.3847/2041-8213/aa8400}

\bibitem[{{Beniamini} {et~al.}(2020){Beniamini}, {Wadiasingh}, \&
  {Metzger}}]{beniamini2020}
{Beniamini}, P., {Wadiasingh}, Z., \& {Metzger}, B.~D. 2020, \mnras, 496, 3390,
  \dodoi{10.1093/mnras/staa1783}

\bibitem[{{Bhat} {et~al.}(2004){Bhat}, {Cordes}, {Camilo}, {Nice}, \&
  {Lorimer}}]{bcc+04}
{Bhat}, N.~D.~R., {Cordes}, J.~M., {Camilo}, F., {Nice}, D.~J., \& {Lorimer},
  D.~R. 2004, \apj, 605, 759, \dodoi{10.1086/382680}

\bibitem[{{Bilous} {et~al.}(2016){Bilous}, {Kondratiev}, {Kramer}, {Keane},
  {Hessels}, {Stappers}, {Malofeev}, {Sobey}, {Breton}, {Cooper}, {Falcke},
  {Karastergiou}, {Michilli}, {Os{\l}owski}, {Sanidas}, {ter Veen}, {van
  Leeuwen}, {Verbiest}, {Weltevrede}, {Zarka}, {Grie{\ss}meier}, {Serylak},
  {Bell}, {Broderick}, {Eisl{\"o}ffel}, {Markoff}, \& {Rowlinson}}]{bkk+16}
{Bilous}, A.~V., {Kondratiev}, V.~I., {Kramer}, M., {et~al.} 2016, \aap, 591,
  A134, \dodoi{10.1051/0004-6361/201527702}

\bibitem[{{Bochenek} {et~al.}(2020){Bochenek}, {Ravi}, {Belov}, {Hallinan},
  {Kocz}, {Kulkarni}, \& {McKenna}}]{sgr_stare}
{Bochenek}, C.~D., {Ravi}, V., {Belov}, K.~V., {et~al.} 2020, \nat, 587, 59,
  \dodoi{10.1038/s41586-020-2872-x}

\bibitem[{{Bondonneau} {et~al.}(2020){Bondonneau}, {Grie{\ss}meier},
  {Theureau}, {Cognard}, {Brionne}, {Kondratiev}, {Bilous}, {McKee}, {Zarka},
  {Viou}, {Guillemot}, {Chen}, {Main}, {Pilia}, {Possenti}, {Serylak},
  {Shaifullah}, {Tiburzi}, {Verbiest}, {Wu}, {Wucknitz}, {Yerin}, {Briand},
  {Cecconi}, {Corbel}, {Dallier}, {Loh}, {Martin}, {Girard}, \&
  {Tasse}}]{bondonneau2020}
{Bondonneau}, L., {Grie{\ss}meier}, J.~M., {Theureau}, G., {et~al.} 2020, arXiv
  e-prints, arXiv:2009.02076.
\newblock \doarXiv{2009.02076}

\bibitem[{{Brentjens} \& {de Bruyn}(2005)}]{2005A&A...441.1217B}
{Brentjens}, M.~A., \& {de Bruyn}, A.~G. 2005, \aap, 441, 1217,
  \dodoi{10.1051/0004-6361:20052990}

\bibitem[{{Broekema} {et~al.}(2018){Broekema}, {Mol}, {Nijboer}, {van
  Amesfoort}, {Brentjens}, {Loose}, {Klijn}, \& {Romein}}]{bmn+18}
{Broekema}, P.~C., {Mol}, J. J.~D., {Nijboer}, R., {et~al.} 2018, Astronomy and
  Computing, 23, 180, \dodoi{10.1016/j.ascom.2018.04.006}

\bibitem[{{Burn}(1966)}]{1966MNRAS.133...67B}
{Burn}, B.~J. 1966, \mnras, 133, 67, \dodoi{10.1093/mnras/133.1.67}

\bibitem[{{Caleb} {et~al.}(2020){Caleb}, {Stappers}, {Abbott}, {Barr},
  {Bezuidenhout}, {Buchner}, {Burgay}, {Chen}, {Cognard}, {Driessen}, {Fender},
  {Hilmarsson}, {Hoang}, {Horn}, {Jankowski}, {Kramer}, {Lorimer}, {Malenta},
  {Morello}, {Pilia}, {Platts}, {Possenti}, {Rajwade}, {Ridolfi}, {Rhodes},
  {Sanidas}, {Serylak}, {Spitler}, {Townsend}, {Weltman}, {Woudt}, \&
  {Wu}}]{caleb2020}
{Caleb}, M., {Stappers}, B.~W., {Abbott}, T.~D., {et~al.} 2020, \mnras, 496,
  4565, \dodoi{10.1093/mnras/staa1791}

\bibitem[{{Chawla} {et~al.}(2020){Chawla}, {Andersen}, {Bhardwaj}, {Fonseca},
  {Josephy}, {Kaspi}, {Michilli}, {Pleunis}, {Bandura}, {Bassa}, {Boyle},
  {Brar}, {Cassanelli}, {Cubranic}, {Dobbs}, {Dong}, {Gaensler}, {Good},
  {Hessels}, {Land ecker}, {Leung}, {Li}, {Lin}, {Masui}, {Mckinven},
  {Mena-Parra}, {Merryfield}, {Meyers}, {Naidu}, {Ng}, {Patel},
  {Rafiei-Ravandi}, {Rahman}, {Sanghavi}, {Scholz}, {Shin}, {Smith}, {Stairs},
  {Tendulkar}, \& {Vanderlinde}}]{chawla2020}
{Chawla}, P., {Andersen}, B.~C., {Bhardwaj}, M., {et~al.} 2020, \apjl, 896,
  L41, \dodoi{10.3847/2041-8213/ab96bf}

\bibitem[{{CHIME/FRB Collaboration}(2018)}]{chi18_overview}
{CHIME/FRB Collaboration}. 2018, \apj, 863, 48,
  \dodoi{10.3847/1538-4357/aad188}

\bibitem[{{CHIME/FRB Collaboration} {et~al.}(2019{\natexlab{a}}){CHIME/FRB
  Collaboration}, {Andersen}, {Bandura}, {Bhardwaj}, {Boubel}, {Boyce},
  {Boyle}, {Brar}, {Cassanelli}, {Chawla}, {Cubranic}, {Deng}, {Dobbs},
  {Fandino}, {Fonseca}, {Gaensler}, {Gilbert}, {Giri}, {Good}, {Halpern},
  {Hill}, {Hinshaw}, {H{\"o}fer}, {Josephy}, {Kaspi}, {Kothes}, {Landecker},
  {Lang}, {Li}, {Lin}, {Masui}, {Mena-Parra}, {Merryfield}, {Mckinven},
  {Michilli}, {Milutinovic}, {Naidu}, {Newburgh}, {Ng}, {Patel}, {Pen},
  {Pinsonneault-Marotte}, {Pleunis}, {Rafiei-Ravandi}, {Rahman}, {Ransom},
  {Renard}, {Scholz}, {Siegel}, {Singh}, {Smith}, {Stairs}, {Tendulkar},
  {Tretyakov}, {Vanderlinde}, {Yadav}, \& {Zwaniga}}]{chi19_8repeaters}
{CHIME/FRB Collaboration}, {Andersen}, B.~C., {Bandura}, K., {et~al.}
  2019{\natexlab{a}}, \apjl, 885, L24, \dodoi{10.3847/2041-8213/ab4a80}

\bibitem[{{CHIME/FRB Collaboration} {et~al.}(2019{\natexlab{b}}){CHIME/FRB
  Collaboration}, {Amiri}, {Bandura}, {Bhardwaj}, {Boubel}, {Boyce}, {Boyle},
  {. Brar}, {Burhanpurkar}, {Cassanelli}, {Chawla}, {Cliche}, {Cubranic},
  {Deng}, {Denman}, {Dobbs}, {Fandino}, {Fonseca}, {Gaensler}, {Gilbert},
  {Gill}, {Giri}, {Good}, {Halpern}, {Hanna}, {Hill}, {Hinshaw}, {H{\"o}fer},
  {Josephy}, {Kaspi}, {Landecker}, {Lang}, {Lin}, {Masui}, {Mckinven},
  {Mena-Parra}, {Merryfield}, {Michilli}, {Milutinovic}, {Moatti}, {Naidu},
  {Newburgh}, {Ng}, {Patel}, {Pen}, {Pinsonneault-Marotte}, {Pleunis},
  {Rafiei-Ravandi}, {Rahman}, {Ransom}, {Renard}, {Scholz}, {Shaw}, {Siegel},
  {Smith}, {Stairs}, {Tendulkar}, {Tretyakov}, {Vanderlinde}, \&
  {Yadav}}]{chi19_r2}
{CHIME/FRB Collaboration}, {Amiri}, M., {Bandura}, K., {et~al.}
  2019{\natexlab{b}}, \nat, 566, 235, \dodoi{10.1038/s41586-018-0864-x}

\bibitem[{{CHIME/FRB Collaboration} {et~al.}(2020{\natexlab{a}}){CHIME/FRB
  Collaboration}, {Amiri}, {Andersen}, {Band ura}, {Bhardwaj}, {Boyle}, {Brar},
  {Chawla}, {Chen}, {Cliche}, {Cubranic}, {Deng}, {Denman}, {Dobbs}, {Dong},
  {Fand ino}, {Fonseca}, {Gaensler}, {Giri}, {Good}, {Halpern}, {Hessels},
  {Hill}, {H{\"o}fer}, {Josephy}, {Kania}, {Karuppusamy}, {Kaspi}, {Keimpema},
  {Kirsten}, {Landecker}, {Lang}, {Leung}, {Li}, {Lin}, {Marcote}, {Masui},
  {McKinven}, {Mena-Parra}, {Merryfield}, {Michilli}, {Milutinovic},
  {Mirhosseini}, {Naidu}, {Newburgh}, {Ng}, {Nimmo}, {Paragi}, {Patel}, {Pen},
  {Pinsonneault-Marotte}, {Pleunis}, {Rafiei-Ravandi}, {Rahman}, {Ransom},
  {Renard}, {Sanghavi}, {Scholz}, {Shaw}, {Shin}, {Siegel}, {Singh}, {Smegal},
  {Smith}, {Stairs}, {Tendulkar}, {Tretyakov}, {Vanderlinde}, {Wang}, {Wang},
  {Wulf}, {Yadav}, \& {Zwaniga}}]{r3_period2020}
{CHIME/FRB Collaboration}, {Amiri}, M., {Andersen}, B.~C., {et~al.}
  2020{\natexlab{a}}, \nat, 582, 351, \dodoi{10.1038/s41586-020-2398-2}

\bibitem[{{CHIME/FRB Collaboration} {et~al.}(2020{\natexlab{b}}){CHIME/FRB
  Collaboration}, {Bandura}, {Bhardwaj}, {Bij}, {Boyce}, {Boyle}, {Brar},
  {Cassanelli}, {Chawla}, {Chen}, {Cliche}, {Cook}, {Cubranic}, {Curtin},
  {Denman}, {Dobbs}, {Dong}, {Fandino}, {Fonseca}, {Gaensler}, {Giri}, {Good},
  {Halpern}, {Hill}, {Hinshaw}, {H{\"o}fer}, {Josephy}, {Kania}, {Kaspi},
  {Landecker}, {Leung}, {Li}, {Lin}, {Masui}, {McKinven}, {Mena-Parra},
  {Merryfield}, {Meyers}, {Michilli}, {Milutinovic}, {Mirhosseini},
  {M{\"u}nchmeyer}, {Naidu}, {Newburgh}, {Ng}, {Patel}, {Pen},
  {Pinsonneault-Marotte}, {Pleunis}, {Quine}, {Rafiei-Ravandi}, {Rahman},
  {Ransom}, {Renard}, {Sanghavi}, {Scholz}, {Shaw}, {Shin}, {Siegel}, {Singh},
  {Smegal}, {Smith}, {Stairs}, {Tan}, {Tendulkar}, {Tretyakov}, {Vanderlinde},
  {Wang}, {Wulf}, \& {Zwaniga}}]{sgr_chime}
{CHIME/FRB Collaboration}, Andersen, B.~C., {Bandura}, K.~M., {Bhardwaj}, M.,
  {et~al.} 2020{\natexlab{b}}, \nat, 587, 54, \dodoi{10.1038/s41586-020-2863-y}

\bibitem[{{Cho} {et~al.}(2020){Cho}, {Macquart}, {Shannon}, {Deller},
  {Morrison}, {Ekers}, {Bannister}, {Farah}, {Qiu}, {Sammons}, {Bailes},
  {Bhandari}, {Day}, {James}, {Phillips}, {Prochaska}, \& {Tuthill}}]{cho2020}
{Cho}, H., {Macquart}, J.-P., {Shannon}, R.~M., {et~al.} 2020, \apjl, 891, L38,
  \dodoi{10.3847/2041-8213/ab7824}

\bibitem[{{Coenen} {et~al.}(2014){Coenen}, {van Leeuwen}, {Hessels},
  {Stappers}, {Kondratiev}, {Alexov}, {Breton}, {Bilous}, {Cooper}, {Falcke},
  {Fallows}, {Gajjar}, {Grie{\ss}meier}, {Hassall}, {Karastergiou}, {Keane},
  {Kramer}, {Kuniyoshi}, {Noutsos}, {Os{\l}owski}, {Pilia}, {Serylak},
  {Schrijvers}, {Sobey}, {ter Veen}, {Verbiest}, {Weltevrede}, {Wijnholds},
  {Zagkouris}, {van Amesfoort}, {Anderson}, {Asgekar}, {Avruch}, {Bell},
  {Bentum}, {Bernardi}, {Best}, {Bonafede}, {Breitling}, {Broderick},
  {Br{\"u}ggen}, {Butcher}, {Ciardi}, {Corstanje}, {Deller}, {Duscha},
  {Eisl{\"o}ffel}, {Fender}, {Ferrari}, {Frieswijk}, {Garrett}, {de Gasperin},
  {de Geus}, {Gunst}, {Hamaker}, {Heald}, {Hoeft}, {van der Horst}, {Juette},
  {Kuper}, {Law}, {Mann}, {McFadden}, {McKay-Bukowski}, {McKean}, {Munk},
  {Orru}, {Paas}, {Pandey-Pommier}, {Polatidis}, {Reich}, {Renting},
  {R{\"o}ttgering}, {Rowlinson}, {Scaife}, {Schwarz}, {Sluman}, {Smirnov},
  {Swinbank}, {Tagger}, {Tang}, {Tasse}, {Thoudam}, {Toribio}, {Vermeulen},
  {Vocks}, {van Weeren}, {Wucknitz}, {Zarka}, \& {Zensus}}]{coe14}
{Coenen}, T., {van Leeuwen}, J., {Hessels}, J.~W.~T., {et~al.} 2014, \aap, 570,
  A60, \dodoi{10.1051/0004-6361/201424495}

\bibitem[{{Coleiro} \& {Chaty}(2013)}]{coleiro2013}
{Coleiro}, A., \& {Chaty}, S. 2013, \apj, 764, 185,
  \dodoi{10.1088/0004-637X/764/2/185}

\bibitem[{{Colgate} \& {Noerdlinger}(1971)}]{colgate1971}
{Colgate}, S.~A., \& {Noerdlinger}, P.~D. 1971, \apj, 165, 509,
  \dodoi{10.1086/150918}

\bibitem[{{Condon} \& {Ransom}(2016)}]{condon2016essential}
{Condon}, J.~J., \& {Ransom}, S.~M. 2016, {Essential Radio Astronomy}
  ({Princeton University Press})

\bibitem[{{Cordes} \& {Chatterjee}(2019)}]{cordes2019}
{Cordes}, J.~M., \& {Chatterjee}, S. 2019, \araa, 57, 417,
  \dodoi{10.1146/annurev-astro-091918-104501}

\bibitem[{{Cruces} {et~al.}(2021){Cruces}, {Spitler}, {Scholz}, {Lynch},
  {Seymour}, {Hessels}, {Gouiff{\'e}s}, {Hilmarsson}, {Kramer}, \&
  {Munjal}}]{cruces2021}
{Cruces}, M., {Spitler}, L.~G., {Scholz}, P., {et~al.} 2021, \mnras, 500, 448,
  \dodoi{10.1093/mnras/staa3223}

\bibitem[{{Dai} {et~al.}(2015){Dai}, {Hobbs}, {Manchester}, {Kerr}, {Shannon},
  {van Straten}, {Mata}, {Bailes}, {Bhat}, {Burke-Spolaor}, {Coles},
  {Johnston}, {Keith}, {Levin}, {Os{\l}owski}, {Reardon}, {Ravi}, {Sarkissian},
  {Tiburzi}, {Toomey}, {Wang}, {Wang}, {Wen}, {Xu}, {Yan}, \& {Zhu}}]{dai2015}
{Dai}, S., {Hobbs}, G., {Manchester}, R.~N., {et~al.} 2015, \mnras, 449, 3223,
  \dodoi{10.1093/mnras/stv508}

\bibitem[{{Day} {et~al.}(2020){Day}, {Deller}, {Shannon}, {Qiu}, {Bannister},
  {Bhandari}, {Ekers}, {Flynn}, {James}, {Macquart}, {Mahony}, {Phillips}, \&
  {Xavier Prochaska}}]{day2020}
{Day}, C.~K., {Deller}, A.~T., {Shannon}, R.~M., {et~al.} 2020, \mnras, 497,
  3335, \dodoi{10.1093/mnras/staa2138}

\bibitem[{{Driessen} {et~al.}(2019){Driessen}, {Janssen}, {Bassa}, {Stappers},
  \& {Stinebring}}]{driessen2019}
{Driessen}, L.~N., {Janssen}, G.~H., {Bassa}, C.~G., {Stappers}, B.~W., \&
  {Stinebring}, D.~R. 2019, \mnras, 483, 1224, \dodoi{10.1093/mnras/sty3192}

\bibitem[{{Du} {et~al.}(2021){Du}, {Wang}, {Wu}, \& {Xu}}]{du2021}
{Du}, S., {Wang}, W., {Wu}, X., \& {Xu}, R. 2021, \mnras, 500, 4678,
  \dodoi{10.1093/mnras/staa3527}

\bibitem[{{Farah} {et~al.}(2018){Farah}, {Flynn}, {Bailes}, {Jameson},
  {Bannister}, {Barr}, {Bateman}, {Bhandari}, {Caleb}, {Campbell-Wilson},
  {Chang}, {Deller}, {Green}, {Hunstead}, {Jankowski}, {Keane}, {Macquart},
  {M{\"o}ller}, {Onken}, {Os{\l}owski}, {Parthasarathy}, {Plant}, {Ravi},
  {Shannon}, {Tucker}, {Venkatraman Krishnan}, \& {Wolf}}]{farah18}
{Farah}, W., {Flynn}, C., {Bailes}, M., {et~al.} 2018, \mnras, 478, 1209,
  \dodoi{10.1093/mnras/sty1122}

\bibitem[{{Fedorova} \& {Rodin}(2019{\natexlab{a}})}]{fa19a}
{Fedorova}, V.~A., \& {Rodin}, A.~E. 2019{\natexlab{a}}, Astronomy Reports, 63,
  39, \dodoi{10.1134/S1063772919010037}

\bibitem[{{Fedorova} \& {Rodin}(2019{\natexlab{b}})}]{fa19b}
---. 2019{\natexlab{b}}, Astronomy Reports, 63, 877,
  \dodoi{10.1134/S1063772919110039}

\bibitem[{{Fonseca} {et~al.}(2020){Fonseca}, {Andersen}, {Bhardwaj}, {Chawla},
  {Good}, {Josephy}, {Kaspi}, {Masui}, {Mckinven}, {Michilli}, {Pleunis},
  {Shin}, {Tendulkar}, {Bandura}, {Boyle}, {Brar}, {Cassanelli}, {Cubranic},
  {Dobbs}, {Dong}, {Gaensler}, {Hinshaw}, {Land ecker}, {Leung}, {Li}, {Lin},
  {Mena-Parra}, {Merryfield}, {Naidu}, {Ng}, {Patel}, {Pen}, {Rafiei-Ravandi},
  {Rahman}, {Ransom}, {Scholz}, {Smith}, {Stairs}, {Vanderlinde}, {Yadav}, \&
  {Zwaniga}}]{fon20}
{Fonseca}, E., {Andersen}, B.~C., {Bhardwaj}, M., {et~al.} 2020, \apjl, 891,
  L6, \dodoi{10.3847/2041-8213/ab7208}

\bibitem[{{Foreman-Mackey} {et~al.}(2013){Foreman-Mackey}, {Hogg}, {Lang}, \&
  {Goodman}}]{fhd+13}
{Foreman-Mackey}, D., {Hogg}, D.~W., {Lang}, D., \& {Goodman}, J. 2013, \pasp,
  125, 306, \dodoi{10.1086/670067}

\bibitem[{{Gajjar} {et~al.}(2018){Gajjar}, {Siemion}, {Price}, {Law},
  {Michilli}, {Hessels}, {Chatterjee}, {Archibald}, {Bower}, {Brinkman},
  {Burke-Spolaor}, {Cordes}, {Croft}, {Enriquez}, {Foster}, {Gizani},
  {Hellbourg}, {Isaacson}, {Kaspi}, {Lazio}, {Lebofsky}, {Lynch}, {MacMahon},
  {McLaughlin}, {Ransom}, {Scholz}, {Seymour}, {Spitler}, {Tendulkar},
  {Werthimer}, \& {Zhang}}]{gajjar2018}
{Gajjar}, V., {Siemion}, A.~P.~V., {Price}, D.~C., {et~al.} 2018, \apj, 863, 2,
  \dodoi{10.3847/1538-4357/aad005}

\bibitem[{{Geyer} {et~al.}(2017){Geyer}, {Karastergiou}, {Kondratiev},
  {Zagkouris}, {Kramer}, {Stappers}, {Grie{\ss}meier}, {Hessels}, {Michilli},
  {Pilia}, \& {Sobey}}]{gkk+17}
{Geyer}, M., {Karastergiou}, A., {Kondratiev}, V.~I., {et~al.} 2017, \mnras,
  470, 2659, \dodoi{10.1093/mnras/stx1151}

\bibitem[{{Gourdji} {et~al.}(2019){Gourdji}, {Michilli}, {Spitler}, {Hessels},
  {Seymour}, {Cordes}, \& {Chatterjee}}]{gourdji2019}
{Gourdji}, K., {Michilli}, D., {Spitler}, L.~G., {et~al.} 2019, \apjl, 877,
  L19, \dodoi{10.3847/2041-8213/ab1f8a}

\bibitem[{{Gruzinov} \& {Levin}(2019)}]{gruzinov2019}
{Gruzinov}, A., \& {Levin}, Y. 2019, \apj, 876, 74,
  \dodoi{10.3847/1538-4357/ab0fa3}

\bibitem[{{Gupta} {et~al.}(2017){Gupta}, {Ajithkumar}, {Kale}, {Nayak},
  {Sabhapathy}, {Sureshkumar}, {Swami}, {Chengalur}, {Ghosh},
  {Ishwara-Chandra}, {Joshi}, {Kanekar}, {Lal}, \& {Roy}}]{gupta2017}
{Gupta}, Y., {Ajithkumar}, B., {Kale}, H.~S., {et~al.} 2017, Current Science,
  113, 707

\bibitem[{{Hardy} {et~al.}(2017){Hardy}, {Dhillon}, {Spitler}, {Littlefair},
  {Ashley}, {De Cia}, {Green}, {Jaroenjittichai}, {Keane}, {Kerry}, {Kramer},
  {Malesani}, {Marsh}, {Parsons}, {Possenti}, {Rattanasoon}, \&
  {Sahman}}]{hardy2017}
{Hardy}, L.~K., {Dhillon}, V.~S., {Spitler}, L.~G., {et~al.} 2017, \mnras, 472,
  2800, \dodoi{10.1093/mnras/stx2153}

\bibitem[{Harris {et~al.}(2020)Harris, Millman, van~der Walt, Gommers,
  Virtanen, Cournapeau, Wieser, Taylor, Berg, Smith, Kern, Picus, Hoyer, van
  Kerkwijk, Brett, Haldane, del R{'{\i}}o, Wiebe, Peterson,
  G{'{e}}rard-Marchant, Sheppard, Reddy, Weckesser, Abbasi, Gohlke, \&
  Oliphant}]{numpy}
Harris, C.~R., Millman, K.~J., van~der Walt, S.~J., {et~al.} 2020, Nature, 585,
  357, \dodoi{10.1038/s41586-020-2649-2}

\bibitem[{{Hassall} {et~al.}(2012){Hassall}, {Stappers}, {Hessels}, {Kramer},
  {Alexov}, {Anderson}, {Coenen}, {Karastergiou}, {Keane}, {Kondratiev},
  {Lazaridis}, {van Leeuwen}, {Noutsos}, {Serylak}, {Sobey}, {Verbiest},
  {Weltevrede}, {Zagkouris}, {Fender}, {Wijers}, {B{\"a}hren}, {Bell},
  {Broderick}, {Corbel}, {Daw}, {Dhillon}, {Eisl{\"o}ffel}, {Falcke},
  {Grie{\ss}meier}, {Jonker}, {Law}, {Markoff}, {Miller-Jones}, {Osten}, {Rol},
  {Scaife}, {Scheers}, {Schellart}, {Spreeuw}, {Swinbank}, {ter Veen}, {Wise},
  {Wijnands}, {Wucknitz}, {Zarka}, {Asgekar}, {Bell}, {Bentum}, {Bernardi},
  {Best}, {Bonafede}, {Boonstra}, {Brentjens}, {Brouw}, {Br{\"u}ggen},
  {Butcher}, {Ciardi}, {Garrett}, {Gerbers}, {Gunst}, {van Haarlem}, {Heald},
  {Hoeft}, {Holties}, {de Jong}, {Koopmans}, {Kuniyoshi}, {Kuper}, {Loose},
  {Maat}, {Masters}, {McKean}, {Meulman}, {Mevius}, {Munk}, {Noordam},
  {Orr{\'u}}, {Paas}, {Pandey-Pommier}, {Pandey}, {Pizzo}, {Polatidis},
  {Reich}, {R{\"o}ttgering}, {Sluman}, {Steinmetz}, {Sterks}, {Tagger}, {Tang},
  {Tasse}, {Vermeulen}, {van Weeren}, {Wijnholds}, \&
  {Yatawatta}}]{hassall2012}
{Hassall}, T.~E., {Stappers}, B.~W., {Hessels}, J.~W.~T., {et~al.} 2012, \aap,
  543, A66, \dodoi{10.1051/0004-6361/201218970}

\bibitem[{{Heald}(2009)}]{2009IAUS..259..591H}
{Heald}, G. 2009, in IAU Symposium, Vol. 259, Cosmic Magnetic Fields: From
  Planets, to Stars and Galaxies, ed. K.~G. {Strassmeier}, A.~G. {Kosovichev},
  \& J.~E. {Beckman}, 591--602, \dodoi{10.1017/S1743921309031421}

\bibitem[{{Hessels} {et~al.}(2019){Hessels}, {Spitler}, {Seymour}, {Cordes},
  {Michilli}, {Lynch}, {Gourdji}, {Archibald}, {Bassa}, {Bower}, {Chatterjee},
  {Connor}, {Crawford}, {Deneva}, {Gajjar}, {Kaspi}, {Keimpema}, {Law},
  {Marcote}, {McLaughlin}, {Paragi}, {Petroff}, {Ransom}, {Scholz}, {Stappers},
  \& {Tendulkar}}]{hessels19}
{Hessels}, J.~W.~T., {Spitler}, L.~G., {Seymour}, A.~D., {et~al.} 2019, \apjl,
  876, L23, \dodoi{10.3847/2041-8213/ab13ae}

\bibitem[{{Hewish} {et~al.}(1968){Hewish}, {Bell}, {Pilkington}, {Scott}, \&
  {Collins}}]{hewish1968}
{Hewish}, A., {Bell}, S.~J., {Pilkington}, J.~D.~H., {Scott}, P.~F., \&
  {Collins}, R.~A. 1968, \nat, 217, 709, \dodoi{10.1038/217709a0}

\bibitem[{{Hilmarsson} {et~al.}(2020){Hilmarsson}, {Michilli}, {Spitler},
  {Wharton}, {Demorest}, {Desvignes}, {Gourdji}, {Hackstein}, {Hessels},
  {Nimmo}, {Seymour}, {Kramer}, \& {McKinven}}]{hilmarsson2020}
{Hilmarsson}, G.~H., {Michilli}, D., {Spitler}, L.~G., {et~al.} 2020, arXiv
  e-prints, arXiv:2009.12135.
\newblock \doarXiv{2009.12135}

\bibitem[{{Hotan} {et~al.}(2004){Hotan}, {van Straten}, \&
  {Manchester}}]{psrchive}
{Hotan}, A.~W., {van Straten}, W., \& {Manchester}, R.~N. 2004, \pasa, 21, 302,
  \dodoi{10.1071/AS04022}

\bibitem[{{Houben} {et~al.}(2019){Houben}, {Spitler}, {ter Veen}, {Rachen},
  {Falcke}, \& {Kramer}}]{hou19}
{Houben}, L.~J.~M., {Spitler}, L.~G., {ter Veen}, S., {et~al.} 2019, \aap, 623,
  A42, \dodoi{10.1051/0004-6361/201833875}

\bibitem[{Hunter(2007)}]{matplotlib}
Hunter, J.~D. 2007, Computing in Science \& Engineering, 9, 90,
  \dodoi{10.1109/MCSE.2007.55}

\bibitem[{{Ilie} {et~al.}(2019){Ilie}, {Johnston}, \& {Weltevrede}}]{ilie2019}
{Ilie}, C.~D., {Johnston}, S., \& {Weltevrede}, P. 2019, \mnras, 483, 2778,
  \dodoi{10.1093/mnras/sty3315}

\bibitem[{{Ioka} \& {Zhang}(2020)}]{iokazhang2020}
{Ioka}, K., \& {Zhang}, B. 2020, \apjl, 893, L26,
  \dodoi{10.3847/2041-8213/ab83fb}

\bibitem[{{Johnston} {et~al.}(2005){Johnston}, {Ball}, {Wang}, \&
  {Manchester}}]{jbwm05}
{Johnston}, S., {Ball}, L., {Wang}, N., \& {Manchester}, R.~N. 2005, \mnras,
  358, 1069, \dodoi{10.1111/j.1365-2966.2005.08854.x}

\bibitem[{{Johnston} {et~al.}(1992){Johnston}, {Manchester}, {Lyne}, {Bailes},
  {Kaspi}, {Qiao}, \& {D'Amico}}]{jml+92}
{Johnston}, S., {Manchester}, R.~N., {Lyne}, A.~G., {et~al.} 1992, \apjl, 387,
  L37, \dodoi{10.1086/186300}

\bibitem[{{Johnston} {et~al.}(1996){Johnston}, {Manchester}, {Lyne}, {D'Amico},
  {Bailes}, {Gaensler}, \& {Nicastro}}]{jml+96}
---. 1996, \mnras, 279, 1026, \dodoi{10.1093/mnras/279.3.1026}

\bibitem[{{Josephy} {et~al.}(2019){Josephy}, {Chawla}, {Fonseca}, {Ng},
  {Patel}, {Pleunis}, {Scholz}, {Andersen}, {Bandura}, {Bhardwaj}, {Boyce},
  {Boyle}, {Brar}, {Cubranic}, {Dobbs}, {Gaensler}, {Gill}, {Giri}, {Good},
  {Halpern}, {Hinshaw}, {Kaspi}, {Landecker}, {Lang}, {Lin}, {Masui},
  {Mckinven}, {Mena-Parra}, {Merryfield}, {Michilli}, {Milutinovic}, {Naidu},
  {Pen}, {Rafiei-Ravandi}, {Rahman}, {Ransom}, {Renard}, {Siegel}, {Smith},
  {Stairs}, {Tendulkar}, {Vanderlinde}, {Yadav}, \& {Zwaniga}}]{josephy2019}
{Josephy}, A., {Chawla}, P., {Fonseca}, E., {et~al.} 2019, \apjl, 882, L18,
  \dodoi{10.3847/2041-8213/ab2c00}

\bibitem[{{Karastergiou} {et~al.}(2015){Karastergiou}, {Chennamangalam},
  {Armour}, {Williams}, {Mort}, {Dulwich}, {Salvini}, {Magro}, {Roberts},
  {Serylak}, {Doo}, {Bilous}, {Breton}, {Falcke}, {Grie{\ss}meier}, {Hessels},
  {Keane}, {Kondratiev}, {Kramer}, {van Leeuwen}, {Noutsos}, {Os{\l}owski},
  {Sobey}, {Stappers}, \& {Weltevrede}}]{kar15}
{Karastergiou}, A., {Chennamangalam}, J., {Armour}, W., {et~al.} 2015, \mnras,
  452, 1254, \dodoi{10.1093/mnras/stv1306}

\bibitem[{{Kaspi} {et~al.}(1996){Kaspi}, {Tauris}, \& {Manchester}}]{ktm96}
{Kaspi}, V.~M., {Tauris}, T.~M., \& {Manchester}, R.~N. 1996, \apj, 459, 717,
  \dodoi{10.1086/176936}

\bibitem[{{Kirsten} {et~al.}(2020){Kirsten}, {Snelders}, {Jenkins}, {Nimmo},
  {van den Eijnden}, {Hessels}, {Gawro{\'n}ski}, \& {Yang}}]{kirsten2020}
{Kirsten}, F., {Snelders}, M.~P., {Jenkins}, M., {et~al.} 2020, Nature
  Astronomy, \dodoi{10.1038/s41550-020-01246-3}

\bibitem[{{Kondratiev} {et~al.}(2016){Kondratiev}, {Verbiest}, {Hessels},
  {Bilous}, {Stappers}, {Kramer}, {Keane}, {Noutsos}, {Os{\l}owski}, {Breton},
  {Hassall}, {Alexov}, {Cooper}, {Falcke}, {Grie{\ss}meier}, {Karastergiou},
  {Kuniyoshi}, {Pilia}, {Sobey}, {ter Veen}, {van Leeuwen}, {Weltevrede},
  {Bell}, {Broderick}, {Corbel}, {Eisl{\"o}ffel}, {Markoff}, {Rowlinson},
  {Swinbank}, {Wijers}, {Wijnands}, \& {Zarka}}]{kvh+16}
{Kondratiev}, V.~I., {Verbiest}, J.~P.~W., {Hessels}, J.~W.~T., {et~al.} 2016,
  \aap, 585, A128, \dodoi{10.1051/0004-6361/201527178}

\bibitem[{{Kulkarni}(2020)}]{kul20}
{Kulkarni}, S.~R. 2020, arXiv e-prints, arXiv:2007.02886.
\newblock \doarXiv{2007.02886}

\bibitem[{{Kumar} {et~al.}(2017){Kumar}, {Lu}, \& {Bhattacharya}}]{kumar2017}
{Kumar}, P., {Lu}, W., \& {Bhattacharya}, M. 2017, \mnras, 468, 2726,
  \dodoi{10.1093/mnras/stx665}

\bibitem[{{Kumar} {et~al.}(2021){Kumar}, {Shannon}, {Flynn}, {Os{\l}owski},
  {Bhandari}, {Day}, {Deller}, {Farah}, {Kaczmarek}, {Kerr}, {Phillips},
  {Price}, {Qiu}, \& {Thyagarajan}}]{kumar2021}
{Kumar}, P., {Shannon}, R.~M., {Flynn}, C., {et~al.} 2021, \mnras, 500, 2525,
  \dodoi{10.1093/mnras/staa3436}

\bibitem[{{Law} {et~al.}(2017){Law}, {Abruzzo}, {Bassa}, {Bower},
  {Burke-Spolaor}, {Butler}, {Cantwell}, {Carey}, {Chatterjee}, {Cordes},
  {Demorest}, {Dowell}, {Fender}, {Gourdji}, {Grainge}, {Hessels}, {Hickish},
  {Kaspi}, {Lazio}, {McLaughlin}, {Michilli}, {Mooley}, {Perrott}, {Ransom},
  {Razavi-Ghods}, {Rupen}, {Scaife}, {Scott}, {Scholz}, {Seymour}, {Spitler},
  {Stovall}, {Tendulkar}, {Titterington}, {Wharton}, \& {Williams}}]{law2017}
{Law}, C.~J., {Abruzzo}, M.~W., {Bassa}, C.~G., {et~al.} 2017, \apj, 850, 76,
  \dodoi{10.3847/1538-4357/aa9700}

\bibitem[{{Levin} {et~al.}(2020){Levin}, {Beloborodov}, \&
  {Bransgrove}}]{levin2020}
{Levin}, Y., {Beloborodov}, A.~M., \& {Bransgrove}, A. 2020, \apjl, 895, L30,
  \dodoi{10.3847/2041-8213/ab8c4c}

\bibitem[{{Lorimer} {et~al.}(2007){Lorimer}, {Bailes}, {McLaughlin},
  {Narkevic}, \& {Crawford}}]{lor07}
{Lorimer}, D.~R., {Bailes}, M., {McLaughlin}, M.~A., {Narkevic}, D.~J., \&
  {Crawford}, F. 2007, Science, 318, 777, \dodoi{10.1126/science.1147532}

\bibitem[{{Luo} {et~al.}(2020){Luo}, {Wang}, {Men}, {Zhang}, {Jiang}, {Xu},
  {Wang}, {Lee}, {Han}, {Zhang}, {Caballero}, {Chen}, {Chen}, {Gan}, {Guo},
  {Hao}, {Huang}, {Jiang}, {Li}, {Li}, {Li}, {Luo}, {Pan}, {Pei}, {Qian},
  {Sun}, {Wang}, {Wang}, {Wen}, {Xu}, {Xu}, {Yan}, {Yan}, {Yu}, {Yuan},
  {Zhang}, \& {Zhu}}]{luo2020}
{Luo}, R., {Wang}, B.~J., {Men}, Y.~P., {et~al.} 2020, \nat, 586, 693,
  \dodoi{10.1038/s41586-020-2827-2}

\bibitem[{{Lyutikov}(2020)}]{lyu2020}
{Lyutikov}, M. 2020, \apj, 889, 135, \dodoi{10.3847/1538-4357/ab55de}

\bibitem[{{Lyutikov} {et~al.}(2020){Lyutikov}, {Barkov}, \&
  {Giannios}}]{lbg2020}
{Lyutikov}, M., {Barkov}, M.~V., \& {Giannios}, D. 2020, \apjl, 893, L39,
  \dodoi{10.3847/2041-8213/ab87a4}

\bibitem[{{Lyutikov} \& {Popov}(2020)}]{lp2020}
{Lyutikov}, M., \& {Popov}, S. 2020, arXiv e-prints, arXiv:2005.05093.
\newblock \doarXiv{2005.05093}

\bibitem[{{Madsen} {et~al.}(2012){Madsen}, {Stairs}, {Kramer}, {Camilo},
  {Hobbs}, {Janssen}, {Lyne}, {Manchester}, {Possenti}, \& {Stappers}}]{msk+12}
{Madsen}, E.~C., {Stairs}, I.~H., {Kramer}, M., {et~al.} 2012, \mnras, 425,
  2378, \dodoi{10.1111/j.1365-2966.2012.21691.x}

\bibitem[{{Majid} {et~al.}(2020){Majid}, {Pearlman}, {Nimmo}, {Hessels},
  {Prince}, {Naudet}, {Kocz}, \& {Horiuchi}}]{majid2020}
{Majid}, W.~A., {Pearlman}, A.~B., {Nimmo}, K., {et~al.} 2020, \apjl, 897, L4,
  \dodoi{10.3847/2041-8213/ab9a4a}

\bibitem[{{Manchester} \& {Taylor}(1972)}]{mt72}
{Manchester}, R.~N., \& {Taylor}, J.~H. 1972, \aplett, 10, 67

\bibitem[{{Marcote} {et~al.}(2020){Marcote}, {Nimmo}, {Hessels}, {Tendulkar},
  {Bassa}, {Paragi}, {Keimpema}, {Bhardwaj}, {Karuppusamy}, {Kaspi}, {Law},
  {Michilli}, {Aggarwal}, {Andersen}, {Archibald}, {Bandura}, {Bower}, {Boyle},
  {Brar}, {Burke-Spolaor}, {Butler}, {Cassanelli}, {Chawla}, {Demorest},
  {Dobbs}, {Fonseca}, {Giri}, {Good}, {Gourdji}, {Josephy}, {Kirichenko},
  {Kirsten}, {Landecker}, {Lang}, {Lazio}, {Li}, {Lin}, {Linford}, {Masui},
  {Mena-Parra}, {Naidu}, {Ng}, {Patel}, {Pen}, {Pleunis}, {Rafiei-Ravandi},
  {Rahman}, {Renard}, {Scholz}, {Siegel}, {Smith}, {Stairs}, {Vanderlinde}, \&
  {Zwaniga}}]{marcote2020}
{Marcote}, B., {Nimmo}, K., {Hessels}, J.~W.~T., {et~al.} 2020, \nat, 577, 190,
  \dodoi{10.1038/s41586-019-1866-z}

\bibitem[{{Marthi} {et~al.}(2020){Marthi}, {Gautam}, {Li}, {Lin}, {Main},
  {Naidu}, {Pen}, \& {Wharton}}]{marthi2020}
{Marthi}, V.~R., {Gautam}, T., {Li}, D.~Z., {et~al.} 2020, \mnras, 499, L16,
  \dodoi{10.1093/mnrasl/slaa148}

\bibitem[{{Mereghetti} {et~al.}(2020){Mereghetti}, {Savchenko}, {Ferrigno},
  {G{\"o}tz}, {Rigoselli}, {Tiengo}, {Bazzano}, {Bozzo}, {Coleiro},
  {Courvoisier}, {Doyle}, {Goldwurm}, {Hanlon}, {Jourdain}, {von Kienlin},
  {Lutovinov}, {Martin-Carrillo}, {Molkov}, {Natalucci}, {Onori}, {Panessa},
  {Rodi}, {Rodriguez}, {S{\'a}nchez-Fern{\'a}ndez}, {Sunyaev}, \&
  {Ubertini}}]{mereghetti2020}
{Mereghetti}, S., {Savchenko}, V., {Ferrigno}, C., {et~al.} 2020, \apjl, 898,
  L29, \dodoi{10.3847/2041-8213/aba2cf}

\bibitem[{{Metzger} {et~al.}(2019){Metzger}, {Margalit}, \&
  {Sironi}}]{metzger2019}
{Metzger}, B.~D., {Margalit}, B., \& {Sironi}, L. 2019, \mnras, 485, 4091,
  \dodoi{10.1093/mnras/stz700}

\bibitem[{{Michilli} \& {Hessels}(2018)}]{sps}
{Michilli}, D., \& {Hessels}, J. W.~T. 2018, {SpS: Single-pulse Searcher}.
\newblock \doeprint{1806.013}

\bibitem[{{Michilli} {et~al.}(2018){Michilli}, {Seymour}, {Hessels}, {Spitler},
  {Gajjar}, {Archibald}, {Bower}, {Chatterjee}, {Cordes}, {Gourdji}, {Heald},
  {Kaspi}, {Law}, {Sobey}, {Adams}, {Bassa}, {Bogdanov}, {Brinkman},
  {Demorest}, {Fernandez}, {Hellbourg}, {Lazio}, {Lynch}, {Maddox}, {Marcote},
  {McLaughlin}, {Paragi}, {Ransom}, {Scholz}, {Siemion}, {Tendulkar}, {van
  Rooy}, {Wharton}, \& {Whitlow}}]{michilli18}
{Michilli}, D., {Seymour}, A., {Hessels}, J.~W.~T., {et~al.} 2018, \nat, 553,
  182, \dodoi{10.1038/nature25149}

\bibitem[{{Nimmo} {et~al.}(2020){Nimmo}, {Hessels}, {Keimpema}, {Archibald},
  {Cordes}, {Karuppusamy}, {Kirsten}, {Li}, {Marcote}, \& {Paragi}}]{nimmo2020}
{Nimmo}, K., {Hessels}, J.~W.~T., {Keimpema}, A., {et~al.} 2020, arXiv
  e-prints, arXiv:2010.05800.
\newblock \doarXiv{2010.05800}

\bibitem[{{Noutsos} {et~al.}(2015){Noutsos}, {Sobey}, {Kondratiev},
  {Weltevrede}, {Verbiest}, {Karastergiou}, {Kramer}, {Kuniyoshi}, {Alexov},
  {Breton}, {Bilous}, {Cooper}, {Falcke}, {Grie{\ss}meier}, {Hassall},
  {Hessels}, {Keane}, {Os{\l}owski}, {Pilia}, {Serylak}, {Stappers}, {ter
  Veen}, {van Leeuwen}, {Zagkouris}, {Anderson}, {B{\"a}hren}, {Bell},
  {Broderick}, {Carbone}, {Cendes}, {Coenen}, {Corbel}, {Eisl{\"o}ffel},
  {Fender}, {Garsden}, {Jonker}, {Law}, {Markoff}, {Masters}, {Miller-Jones},
  {Molenaar}, {Osten}, {Pietka}, {Rol}, {Rowlinson}, {Scheers}, {Spreeuw},
  {Staley}, {Stewart}, {Swinbank}, {Wijers}, {Wijnands}, {Wise}, {Zarka}, \&
  {van der Horst}}]{noutsos2015}
{Noutsos}, A., {Sobey}, C., {Kondratiev}, V.~I., {et~al.} 2015, \aap, 576, A62,
  \dodoi{10.1051/0004-6361/201425186}

\bibitem[{{Parent} {et~al.}(2020){Parent}, {Chawla}, {Kaspi}, {Agazie},
  {Blumer}, {DeCesar}, {Fiore}, {Fonseca}, {Hessels}, {Kaplan}, {Kondratiev},
  {LaRose}, {Levin}, {Lewis}, {Lynch}, {McEwen}, {McLaughlin}, {Mingyar}, {Al
  Noori}, {Ransom}, {Roberts}, {Schmiedekamp}, {Schmiedekamp}, {Siemens},
  {Spiewak}, {Stairs}, {Surnis}, {Swiggum}, \& {van Leeuwen}}]{pck+20}
{Parent}, E., {Chawla}, P., {Kaspi}, V.~M., {et~al.} 2020, \apj, 904, 92,
  \dodoi{10.3847/1538-4357/abbdf6}

\bibitem[{{Pastor-Marazuela} {et~al.}(2020){Pastor-Marazuela}, {Connor}, {van
  Leeuwen}, {Maan}, {ter Veen}, {Bilous}, {Oostrum}, {Petroff}, {Straal},
  {Vohl}, {Attema}, {Boersma}, {Kooistra}, {van der Schuur}, {Sclocco},
  {Smits}, {Adams}, {Adebahr}, {de Blok}, {Coolen}, {Damstra}, {D{\'e}nes},
  {Hess}, {van der Hulst}, {Hut}, {Ivashina}, {Kutkin}, {Loose}, {Lucero},
  {Mika}, {Moss}, {Mulder}, {Norden}, {Oosterloo}, {Orr{\'u}}, \&
  {Wijnholds}}]{pcl+20}
{Pastor-Marazuela}, I., {Connor}, L., {van Leeuwen}, J., {et~al.} 2020, {\it
  submitted}

\bibitem[{{Pearlman} {et~al.}(2020){Pearlman}, {Majid}, {Prince}, {Nimmo},
  {Hessels}, {Naudet}, \& {Kocz}}]{pearlman2020}
{Pearlman}, A.~B., {Majid}, W.~A., {Prince}, T.~A., {et~al.} 2020, \apjl, 905,
  L27, \dodoi{10.3847/2041-8213/abca31}

\bibitem[{{Petroff} {et~al.}(2019){Petroff}, {Hessels}, \&
  {Lorimer}}]{petroff2019}
{Petroff}, E., {Hessels}, J.~W.~T., \& {Lorimer}, D.~R. 2019, \aapr, 27, 4,
  \dodoi{10.1007/s00159-019-0116-6}

\bibitem[{{Petroff} {et~al.}(2015){Petroff}, {Johnston}, {Keane}, {van
  Straten}, {Bailes}, {Barr}, {Barsdell}, {Burke-Spolaor}, {Caleb}, {Champion},
  {Flynn}, {Jameson}, {Kramer}, {Ng}, {Possenti}, \& {Stappers}}]{petroff2015}
{Petroff}, E., {Johnston}, S., {Keane}, E.~F., {et~al.} 2015, \mnras, 454, 457,
  \dodoi{10.1093/mnras/stv1953}

\bibitem[{{Phinney} \& {Taylor}(1979)}]{phinney1979}
{Phinney}, S., \& {Taylor}, J.~H. 1979, \nat, 277, 117,
  \dodoi{10.1038/277117a0}

\bibitem[{{Pilia} {et~al.}(2020){Pilia}, {Burgay}, {Possenti}, {Ridolfi},
  {Gajjar}, {Corongiu}, {Perrodin}, {Bernardi}, {Naldi}, {Pupillo},
  {Ambrosino}, {Bianchi}, {Burtovoi}, {Casella}, {Casentini}, {Cecconi},
  {Ferrigno}, {Fiori}, {Gendreau}, {Ghedina}, {Naletto}, {Nicastro}, {Ochner},
  {Palazzi}, {Panessa}, {Papitto}, {Pittori}, {Rea}, {Castillo}, {Savchenko},
  {Setti}, {Tavani}, {Trois}, {Trudu}, {Turatto}, {Ursi}, {Verrecchia}, \&
  {Zampieri}}]{pilia2020}
{Pilia}, M., {Burgay}, M., {Possenti}, A., {et~al.} 2020, \apjl, 896, L40,
  \dodoi{10.3847/2041-8213/ab96c0}

\bibitem[{{Piro}(2016)}]{piro2016}
{Piro}, A.~L. 2016, \apjl, 824, L32, \dodoi{10.3847/2041-8205/824/2/L32}

\bibitem[{{Platts} {et~al.}(2019){Platts}, {Weltman}, {Walters}, {Tendulkar},
  {Gordin}, \& {Kandhai}}]{pla18}
{Platts}, E., {Weltman}, A., {Walters}, A., {et~al.} 2019, \physrep, 821, 1,
  \dodoi{10.1016/j.physrep.2019.06.003}

\bibitem[{{Popov}(2020)}]{popov2020}
{Popov}, S.~B. 2020, Research Notes of the American Astronomical Society, 4,
  98, \dodoi{10.3847/2515-5172/aba0af}

\bibitem[{{Purcell} {et~al.}(2020){Purcell}, {Van Eck}, {West}, {Sun}, \&
  {Gaensler}}]{rmtools}
{Purcell}, C.~R., {Van Eck}, C.~L., {West}, J., {Sun}, X.~H., \& {Gaensler},
  B.~M. 2020, {RM-Tools: Rotation measure (RM) synthesis and Stokes
  QU-fitting}.
\newblock \doeprint{2005.003}

\bibitem[{{Rajwade} {et~al.}(2020){Rajwade}, {Mickaliger}, {Stappers},
  {Morello}, {Agarwal}, {Bassa}, {Breton}, {Caleb}, {Karastergiou}, {Keane}, \&
  {Lorimer}}]{rajwade2020}
{Rajwade}, K.~M., {Mickaliger}, M.~B., {Stappers}, B.~W., {et~al.} 2020,
  \mnras, 495, 3551, \dodoi{10.1093/mnras/staa1237}

\bibitem[{{Ransom}(2001)}]{ran01}
{Ransom}, S.~M. 2001, PhD thesis, Harvard University

\bibitem[{{Ransom} {et~al.}(2002){Ransom}, {Eikenberry}, \&
  {Middleditch}}]{rem02}
{Ransom}, S.~M., {Eikenberry}, S.~S., \& {Middleditch}, J. 2002, \aj, 124,
  1788, \dodoi{10.1086/342285}

\bibitem[{{Reddy} {et~al.}(2017){Reddy}, {Kudale}, {Gokhale}, {Halagalli},
  {Raskar}, {de}, {Gnanaraj}, {Ajith Kumar}, \& {Gupta}}]{reddy17}
{Reddy}, S.~H., {Kudale}, S., {Gokhale}, U., {et~al.} 2017, Journal of
  Astronomical Instrumentation, 6, 1641011, \dodoi{10.1142/S2251171716410117}

\bibitem[{{Rowlinson} {et~al.}(2016){Rowlinson}, {Bell}, {Murphy}, {Trott},
  {Hurley-Walker}, {Johnston}, {Tingay}, {Kaplan}, {Carbone}, {Hancock},
  {Feng}, {Offringa}, {Bernardi}, {Bowman}, {Briggs}, {Cappallo}, {Deshpande},
  {Gaensler}, {Greenhill}, {Hazelton}, {Johnston-Hollitt}, {Lonsdale},
  {McWhirter}, {Mitchell}, {Morales}, {Morgan}, {Oberoi}, {Ord}, {Prabu},
  {Udaya Shankar}, {Srivani}, {Subrahmanyan}, {Wayth}, {Webster}, {Williams},
  \& {Williams}}]{row16}
{Rowlinson}, A., {Bell}, M.~E., {Murphy}, T., {et~al.} 2016, \mnras, 458, 3506,
  \dodoi{10.1093/mnras/stw451}

\bibitem[{{Sana} {et~al.}(2012){Sana}, {de Mink}, {de Koter}, {Langer},
  {Evans}, {Gieles}, {Gosset}, {Izzard}, {Le Bouquin}, \&
  {Schneider}}]{sana2012}
{Sana}, H., {de Mink}, S.~E., {de Koter}, A., {et~al.} 2012, Science, 337, 444,
  \dodoi{10.1126/science.1223344}

\bibitem[{{Sanidas} {et~al.}(2019){Sanidas}, {Cooper}, {Bassa}, {Hessels},
  {Kondratiev}, {Michilli}, {Stappers}, {Tan}, {van Leeuwen}, {Cerrigone},
  {Fallows}, {Iacobelli}, {Orr{\'u}}, {Pizzo}, {Shulevski}, {Toribio}, {ter
  Veen}, {Zucca}, {Bondonneau}, {Grie{\ss}meier}, {Karastergiou}, {Kramer}, \&
  {Sobey}}]{scb+19}
{Sanidas}, S., {Cooper}, S., {Bassa}, C.~G., {et~al.} 2019, \aap, 626, A104,
  \dodoi{10.1051/0004-6361/201935609}

\bibitem[{{Scholz} {et~al.}(2017){Scholz}, {Bogdanov}, {Hessels}, {Lynch},
  {Spitler}, {Bassa}, {Bower}, {Burke-Spolaor}, {Butler}, {Chatterjee},
  {Cordes}, {Gourdji}, {Kaspi}, {Law}, {Marcote}, {McLaughlin}, {Michilli},
  {Paragi}, {Ransom}, {Seymour}, {Tendulkar}, \& {Wharton}}]{scholz17}
{Scholz}, P., {Bogdanov}, S., {Hessels}, J.~W.~T., {et~al.} 2017, \apj, 846,
  80, \dodoi{10.3847/1538-4357/aa8456}

\bibitem[{{Scholz} {et~al.}(2020){Scholz}, {Cook}, {Cruces}, {Hessels},
  {Kaspi}, {Majid}, {Naidu}, {Pearlman}, {Spitler}, {Bandura}, {Bhardwaj},
  {Cassanelli}, {Chawla}, {Gaensler}, {Good}, {Josephy}, {Karuppusamy},
  {Keimpema}, {Kirichenko}, {Kirsten}, {Kocz}, {Leung}, {Marcote}, {Masui},
  {Mena-Parra}, {Merryfield}, {Michilli}, {Naudet}, {Nimmo}, {Pleunis},
  {Prince}, {Rafiei-Ravandi}, {Rahman}, {Shin}, {Smith}, {Stairs}, {Tendulkar},
  \& {Vanderlinde}}]{scholz2020}
{Scholz}, P., {Cook}, A., {Cruces}, M., {et~al.} 2020, \apj, 901, 165,
  \dodoi{10.3847/1538-4357/abb1a8}

\bibitem[{{Seymour} {et~al.}(2019){Seymour}, {Michilli}, \&
  {Pleunis}}]{dmphase}
{Seymour}, A., {Michilli}, D., \& {Pleunis}, Z. 2019, {DM\_phase: Algorithm for
  correcting dispersion of radio signals}.
\newblock \doeprint{1910.004}

\bibitem[{{Shannon} {et~al.}(2018){Shannon}, {Macquart}, {Bannister}, {Ekers},
  {James}, {Os{\l}owski}, {Qiu}, {Sammons}, {Hotan}, {Voronkov}, {Beresford},
  {Brothers}, {Brown}, {Bunton}, {Chippendale}, {Haskins}, {Leach},
  {Marquarding}, {McConnell}, {Pilawa}, {Sadler}, {Troup}, {Tuthill},
  {Whiting}, {Allison}, {Anderson}, {Bell}, {Collier}, {G{\"u}rkan}, {Heald},
  \& {Riseley}}]{shannon2018}
{Shannon}, R.~M., {Macquart}, J.~P., {Bannister}, K.~W., {et~al.} 2018, \nat,
  562, 386, \dodoi{10.1038/s41586-018-0588-y}

\bibitem[{{Sobey} {et~al.}(2019){Sobey}, {Bilous}, {Grie{\ss}meier}, {Hessels},
  {Karastergiou}, {Keane}, {Kondratiev}, {Kramer}, {Michilli}, {Noutsos},
  {Pilia}, {Polzin}, {Stappers}, {Tan}, {van Leeuwen}, {Verbiest},
  {Weltevrede}, {Heald}, {Alves}, {Carretti}, {En{\ss}lin}, {Haverkorn},
  {Iacobelli}, {Reich}, \& {Van Eck}}]{sob19}
{Sobey}, C., {Bilous}, A.~V., {Grie{\ss}meier}, J.~M., {et~al.} 2019, \mnras,
  484, 3646, \dodoi{10.1093/mnras/stz214}

\bibitem[{{Sob'yanin}(2020)}]{sobyanin2020}
{Sob'yanin}, D.~N. 2020, \mnras, 497, 1001, \dodoi{10.1093/mnras/staa1976}

\bibitem[{{Sokolowski} {et~al.}(2018){Sokolowski}, {Bhat}, {Macquart},
  {Shannon}, {Bannister}, {Ekers}, {Scott}, {Beardsley}, {Crosse}, {Emrich},
  {Franzen}, {Gaensler}, {Horsley}, {Johnston-Hollitt}, {Kaplan}, {Kenney},
  {Morales}, {Pallot}, {Sleap}, {Steele}, {Tingay}, {Trott}, {Walker}, {Wayth},
  {Williams}, \& {Wu}}]{sokolowski2018}
{Sokolowski}, M., {Bhat}, N.~D.~R., {Macquart}, J.~P., {et~al.} 2018, \apjl,
  867, L12, \dodoi{10.3847/2041-8213/aae58d}

\bibitem[{{Sotomayor-Beltran} {et~al.}(2013){Sotomayor-Beltran}, {Sobey},
  {Hessels}, {de Bruyn}, {Noutsos}, {Alexov}, {Anderson}, {Asgekar}, {Avruch},
  {Beck}, {Bell}, {Bell}, {Bentum}, {Bernardi}, {Best}, {Birzan}, {Bonafede},
  {Breitling}, {Broderick}, {Brouw}, {Br{\"u}ggen}, {Ciardi}, {de Gasperin},
  {Dettmar}, {van Duin}, {Duscha}, {Eisl{\"o}ffel}, {Falcke}, {Fallows},
  {Fender}, {Ferrari}, {Frieswijk}, {Garrett}, {Grie{\ss}meier}, {Grit},
  {Gunst}, {Hassall}, {Heald}, {Hoeft}, {Horneffer}, {Iacobelli}, {Juette},
  {Karastergiou}, {Keane}, {Kohler}, {Kramer}, {Kondratiev}, {Koopmans},
  {Kuniyoshi}, {Kuper}, {van Leeuwen}, {Maat}, {Macario}, {Markoff}, {McKean},
  {Mulcahy}, {Munk}, {Orru}, {Paas}, {Pandey-Pommier}, {Pilia}, {Pizzo},
  {Polatidis}, {Reich}, {R{\"o}ttgering}, {Serylak}, {Sluman}, {Stappers},
  {Tagger}, {Tang}, {Tasse}, {ter Veen}, {Vermeulen}, {van Weeren}, {Wijers},
  {Wijnholds}, {Wise}, {Wucknitz}, {Yatawatta}, \& {Zarka}}]{sot13}
{Sotomayor-Beltran}, C., {Sobey}, C., {Hessels}, J.~W.~T., {et~al.} 2013, \aap,
  552, A58, \dodoi{10.1051/0004-6361/201220728}

\bibitem[{{Spitler} {et~al.}(2016){Spitler}, {Scholz}, {Hessels}, {Bogdanov},
  {Brazier}, {Camilo}, {Chatterjee}, {Cordes}, {Crawford}, {Deneva}, {Ferdman},
  {Freire}, {Kaspi}, {Lazarus}, {Lynch}, {Madsen}, {McLaughlin}, {Patel},
  {Ransom}, {Seymour}, {Stairs}, {Stappers}, {van Leeuwen}, \& {Zhu}}]{spi16}
{Spitler}, L.~G., {Scholz}, P., {Hessels}, J.~W.~T., {et~al.} 2016, \nat, 531,
  202, \dodoi{10.1038/nature17168}

\bibitem[{{Stappers} {et~al.}(2011){Stappers}, {Hessels}, {Alexov}, {Anderson},
  {Coenen}, {Hassall}, {Karastergiou}, {Kondratiev}, {Kramer}, {van Leeuwen},
  {Mol}, {Noutsos}, {Romein}, {Weltevrede}, {Fender}, {Wijers}, {B{\"a}hren},
  {Bell}, {Broderick}, {Daw}, {Dhillon}, {Eisl{\"o}ffel}, {Falcke},
  {Griessmeier}, {Law}, {Markoff}, {Miller-Jones}, {Scheers}, {Spreeuw},
  {Swinbank}, {Ter Veen}, {Wise}, {Wucknitz}, {Zarka}, {Anderson}, {Asgekar},
  {Avruch}, {Beck}, {Bennema}, {Bentum}, {Best}, {Bregman}, {Brentjens}, {van
  de Brink}, {Broekema}, {Brouw}, {Br{\"u}ggen}, {de Bruyn}, {Butcher},
  {Ciardi}, {Conway}, {Dettmar}, {van Duin}, {van Enst}, {Garrett}, {Gerbers},
  {Grit}, {Gunst}, {van Haarlem}, {Hamaker}, {Heald}, {Hoeft}, {Holties},
  {Horneffer}, {Koopmans}, {Kuper}, {Loose}, {Maat}, {McKay-Bukowski},
  {McKean}, {Miley}, {Morganti}, {Nijboer}, {Noordam}, {Norden}, {Olofsson},
  {Pandey-Pommier}, {Polatidis}, {Reich}, {R{\"o}ttgering}, {Schoenmakers},
  {Sluman}, {Smirnov}, {Steinmetz}, {Sterks}, {Tagger}, {Tang}, {Vermeulen},
  {Vermaas}, {Vogt}, {de Vos}, {Wijnholds}, {Yatawatta}, \& {Zensus}}]{sha+11}
{Stappers}, B.~W., {Hessels}, J.~W.~T., {Alexov}, A., {et~al.} 2011, \aap, 530,
  A80, \dodoi{10.1051/0004-6361/201116681}

\bibitem[{{Susobhanan} {et~al.}(2020){Susobhanan}, {Maan}, {Joshi}, {Prabu},
  {Desai}, {Gupta}, {Gopakumar}, {Dhanda Batra}, {Choudhary}, {Surnis}, {Dey},
  {Singha}, {Nobleson}, {Bagchi}, {Basu}, {Bethapudi}, {De}, {Girgaonkar},
  {Krishnakumar}, {Manoharan}, {Naidu}, {Pathak}, \& {Chaitanya
  Susarla}}]{Susobhanan20}
{Susobhanan}, A., {Maan}, Y., {Joshi}, B.~C., {et~al.} 2020, arXiv e-prints,
  arXiv:2007.02930.
\newblock \doarXiv{2007.02930}

\bibitem[{{Tendulkar} {et~al.}(2020){Tendulkar}, {Gil de Paz}, {Kirichenko},
  {Hessels}, {Bhardwaj}, {{\'A}vila}, {Bassa}, {Chawla}, {Fonseca}, {Kaspi},
  {Keimpema}, {Kirsten}, {Lazio}, {Marcote}, {Masui}, {Nimmo}, {Paragi},
  {Rahman}, {Reverte Pay{\'a}}, {Scholz}, \& {Stairs}}]{tendulkar2020}
{Tendulkar}, S.~P., {Gil de Paz}, A., {Kirichenko}, A.~Y., {et~al.} 2020, arXiv
  e-prints, arXiv:2011.03257.
\newblock \doarXiv{2011.03257}

\bibitem[{{Thornton} {et~al.}(2013){Thornton}, {Stappers}, {Bailes},
  {Barsdell}, {Bates}, {Bhat}, {Burgay}, {Burke-Spolaor}, {Champion}, {Coster},
  {D'Amico}, {Jameson}, {Johnston}, {Keith}, {Kramer}, {Levin}, {Milia}, {Ng},
  {Possenti}, \& {van Straten}}]{tho13}
{Thornton}, D., {Stappers}, B., {Bailes}, M., {et~al.} 2013, Science, 341, 53,
  \dodoi{10.1126/science.1236789}

\bibitem[{{Tingay} {et~al.}(2015){Tingay}, {Trott}, {Wayth}, {Bernardi},
  {Bowman}, {Briggs}, {Cappallo}, {Deshpande}, {Feng}, {Gaensler}, {Greenhill},
  {Hancock}, {Hazelton}, {Johnston-Hollitt}, {Kaplan}, {Lonsdale}, {McWhirter},
  {Mitchell}, {Morales}, {Morgan}, {Murphy}, {Oberoi}, {Prabu}, {Udaya
  Shankar}, {Srivani}, {Subrahmanyan}, {Webster}, {Williams}, \&
  {Williams}}]{tingay2015}
{Tingay}, S.~J., {Trott}, C.~M., {Wayth}, R.~B., {et~al.} 2015, \aj, 150, 199,
  \dodoi{10.1088/0004-6256/150/6/199}

\bibitem[{{Toribio San Cipriano} {et~al.}(2017){Toribio San Cipriano},
  {Dom{\'\i}nguez-Guzm{\'a}n}, {Esteban}, {Garc{\'\i}a-Rojas}, {Mesa-Delgado},
  {Bresolin}, {Rodr{\'\i}guez}, \& {Sim{\'o}n-D{\'\i}az}}]{tde+17}
{Toribio San Cipriano}, L., {Dom{\'\i}nguez-Guzm{\'a}n}, G., {Esteban}, C.,
  {et~al.} 2017, \mnras, 467, 3759, \dodoi{10.1093/mnras/stx328}

\bibitem[{{van Haarlem} {et~al.}(2013){van Haarlem}, {Wise}, {Gunst}, {Heald},
  {McKean}, {Hessels}, {de Bruyn}, {Nijboer}, {Swinbank}, {Fallows},
  {Brentjens}, {Nelles}, {Beck}, {Falcke}, {Fender}, {H{\"o}randel},
  {Koopmans}, {Mann}, {Miley}, {R{\"o}ttgering}, {Stappers}, {Wijers},
  {Zaroubi}, {van den Akker}, {Alexov}, {Anderson}, {Anderson}, {van Ardenne},
  {Arts}, {Asgekar}, {Avruch}, {Batejat}, {B{\"a}hren}, {Bell}, {Bell}, {van
  Bemmel}, {Bennema}, {Bentum}, {Bernardi}, {Best}, {B{\^\i}rzan}, {Bonafede},
  {Boonstra}, {Braun}, {Bregman}, {Breitling}, {van de Brink}, {Broderick},
  {Broekema}, {Brouw}, {Br{\"u}ggen}, {Butcher}, {van Cappellen}, {Ciardi},
  {Coenen}, {Conway}, {Coolen}, {Corstanje}, {Damstra}, {Davies}, {Deller},
  {Dettmar}, {van Diepen}, {Dijkstra}, {Donker}, {Doorduin}, {Dromer}, {Drost},
  {van Duin}, {Eisl{\"o}ffel}, {van Enst}, {Ferrari}, {Frieswijk}, {Gankema},
  {Garrett}, {de Gasperin}, {Gerbers}, {de Geus}, {Grie{\ss}meier}, {Grit},
  {Gruppen}, {Hamaker}, {Hassall}, {Hoeft}, {Holties}, {Horneffer}, {van der
  Horst}, {van Houwelingen}, {Huijgen}, {Iacobelli}, {Intema}, {Jackson},
  {Jelic}, {de Jong}, {Juette}, {Kant}, {Karastergiou}, {Koers}, {Kollen},
  {Kondratiev}, {Kooistra}, {Koopman}, {Koster}, {Kuniyoshi}, {Kramer},
  {Kuper}, {Lambropoulos}, {Law}, {van Leeuwen}, {Lemaitre}, {Loose}, {Maat},
  {Macario}, {Markoff}, {Masters}, {McFadden}, {McKay-Bukowski}, {Meijering},
  {Meulman}, {Mevius}, {Middelberg}, {Millenaar}, {Miller-Jones}, {Mohan},
  {Mol}, {Morawietz}, {Morganti}, {Mulcahy}, {Mulder}, {Munk}, {Nieuwenhuis},
  {van Nieuwpoort}, {Noordam}, {Norden}, {Noutsos}, {Offringa}, {Olofsson},
  {Omar}, {Orr{\'u}}, {Overeem}, {Paas}, {Pandey-Pommier}, {Pandey}, {Pizzo},
  {Polatidis}, {Rafferty}, {Rawlings}, {Reich}, {de Reijer}, {Reitsma},
  {Renting}, {Riemers}, {Rol}, {Romein}, {Roosjen}, {Ruiter}, {Scaife}, {van
  der Schaaf}, {Scheers}, {Schellart}, {Schoenmakers}, {Schoonderbeek},
  {Serylak}, {Shulevski}, {Sluman}, {Smirnov}, {Sobey}, {Spreeuw}, {Steinmetz},
  {Sterks}, {Stiepel}, {Stuurwold}, {Tagger}, {Tang}, {Tasse}, {Thomas},
  {Thoudam}, {Toribio}, {van der Tol}, {Usov}, {van Veelen}, {van der Veen},
  {ter Veen}, {Verbiest}, {Vermeulen}, {Vermaas}, {Vocks}, {Vogt}, {de Vos},
  {van der Wal}, {van Weeren}, {Weggemans}, {Weltevrede}, {White}, {Wijnholds},
  {Wilhelmsson}, {Wucknitz}, {Yatawatta}, {Zarka}, {Zensus}, \& {van
  Zwieten}}]{hwg+13}
{van Haarlem}, M.~P., {Wise}, M.~W., {Gunst}, A.~W., {et~al.} 2013, \aap, 556,
  A2, \dodoi{10.1051/0004-6361/201220873}

\bibitem[{{van Straten} \& {Bailes}(2011)}]{sb10}
{van Straten}, W., \& {Bailes}, M. 2011, \pasa, 28, 1, \dodoi{10.1071/AS10021}

\bibitem[{{Vedantham} \& {Ravi}(2019)}]{vedantham2019}
{Vedantham}, H.~K., \& {Ravi}, V. 2019, \mnras, 485, L78,
  \dodoi{10.1093/mnrasl/slz038}

\bibitem[{Virtanen {et~al.}(2020)Virtanen, Gommers, Oliphant, Haberland, Reddy,
  Cournapeau, Burovski, Peterson, Weckesser, Bright, {van der Walt}, Brett,
  Wilson, Millman, Mayorov, Nelson, Jones, Kern, Larson, Carey, Polat, Feng,
  Moore, {VanderPlas}, Laxalde, Perktold, Cimrman, Henriksen, Quintero, Harris,
  Archibald, Ribeiro, Pedregosa, {van Mulbregt}, \& {SciPy 1.0
  Contributors}}]{scipy}
Virtanen, P., Gommers, R., Oliphant, T.~E., {et~al.} 2020, Nature Methods, 17,
  261, \dodoi{10.1038/s41592-019-0686-2}

\bibitem[{{Walter} {et~al.}(2015){Walter}, {Lutovinov}, {Bozzo}, \&
  {Tsygankov}}]{walter2015}
{Walter}, R., {Lutovinov}, A.~A., {Bozzo}, E., \& {Tsygankov}, S.~S. 2015,
  \aapr, 23, 2, \dodoi{10.1007/s00159-015-0082-6}

\bibitem[{{Wang} {et~al.}(2019){Wang}, {Zhang}, {Chen}, \& {Xu}}]{wang2019}
{Wang}, W., {Zhang}, B., {Chen}, X., \& {Xu}, R. 2019, \apjl, 876, L15,
  \dodoi{10.3847/2041-8213/ab1aab}

\bibitem[{{Xue} {et~al.}(2019){Xue}, {Ord}, {Tremblay}, {Bhat}, {Sobey},
  {Meyers}, {McSweeney}, \& {Swainston}}]{xue19}
{Xue}, M., {Ord}, S.~M., {Tremblay}, S.~E., {et~al.} 2019, \pasa, 36, e025,
  \dodoi{10.1017/pasa.2019.19}

\bibitem[{{Yang} \& {Zou}(2020)}]{yangzou2020}
{Yang}, H., \& {Zou}, Y.-C. 2020, \apjl, 893, L31,
  \dodoi{10.3847/2041-8213/ab800f}

\bibitem[{{Zanazzi} \& {Lai}(2020)}]{zanazzi2020}
{Zanazzi}, J.~J., \& {Lai}, D. 2020, \apjl, 892, L15,
  \dodoi{10.3847/2041-8213/ab7cdd}

\bibitem[{{Zhang}(2017)}]{zhang2017}
{Zhang}, B. 2017, \apjl, 836, L32, \dodoi{10.3847/2041-8213/aa5ded}

\bibitem[{{Zhang} \& {Gao}(2020)}]{zhanggao2020}
{Zhang}, X., \& {Gao}, H. 2020, \mnras, 498, L1, \dodoi{10.1093/mnrasl/slaa116}

\bibitem[{{Zhang} {et~al.}(2018){Zhang}, {Gajjar}, {Foster}, {Siemion},
  {Cordes}, {Law}, \& {Wang}}]{zhang18}
{Zhang}, Y.~G., {Gajjar}, V., {Foster}, G., {et~al.} 2018, \apj, 866, 149,
  \dodoi{10.3847/1538-4357/aadf31}

\end{thebibliography}
\bibliographystyle{aasjournal}

\end{document}